\documentclass[fleqn,usenatbib]{mnras}

\usepackage[T1]{fontenc}

\DeclareRobustCommand{\VAN}[3]{#2}
\let\VANthebibliography\thebibliography
\def\thebibliography{\DeclareRobustCommand{\VAN}[3]{##3}\VANthebibliography}

\usepackage{graphicx}	
\usepackage{amsmath}	
\usepackage{amssymb}
\usepackage{gensymb}

\usepackage{subcaption}
\usepackage{geometry}
\usepackage{float}
\usepackage[font=small,labelfont=bf]{caption}
\usepackage{multicol}
\usepackage{xcolor}

\usepackage{newtxtext,newtxmath}

\newcommand{\GG}[1]{}

\title[A new convection scheme]{A new convection scheme for GCMs of temperate sub-Neptunes}

\author[E. F. L. Barrier et al.]{
Edouard F. L. Barrier,$^{1}$\thanks{E-mail: efxlb2@ast.cam.ac.uk}
Nikku Madhusudhan$^{1}$\thanks{E-mail: nmadhu@ast.cam.ac.uk}
\\
$^{1}$Institute of Astronomy, University of Cambridge, Madingley Road, Cambridge, CB3 0HA, UK
}

\date{Accepted XXX. Received YYY; in original form ZZZ}

\pubyear{2024}

\begin{document}
\label{firstpage}
\pagerange{\pageref{firstpage}--\pageref{lastpage}}
\maketitle

\begin{abstract}
Atmospheric characterisation of temperate sub-Neptunes is the new frontier of exoplanetary science with recent JWST observations of possible Hycean world K2-18~b. Accurate modelling of atmospheric processes is essential to interpreting high-precision spectroscopic data given the wide range of possible conditions in the sub-Neptune regime, including on potentially habitable planets. Notably, convection is an important process which can operate in different modes across sub-Neptune conditions. Convection can act very differently in atmospheres with a high condensible mass fraction (non-dilute atmospheres) or with a lighter background gas, e.g. water convection in a H$_2$-rich atmosphere, and can be much weaker or even shut down entirely in the latter case. We present a new mass-flux scheme which can capture these variations and simulate convection over a wide range of parameter space for use in 3D general circulation models (GCMs). We validate our scheme for two representative cases, a terrestrial-like atmosphere and a mini-Neptune atmosphere. In the terrestrial case, considering TRAPPIST-1e with an Earth-like atmosphere, the model performs near-identically to Earth-tuned models in an Earth-like convection case. In the mini-Neptune case, considering the bulk properties of K2-18~b and assuming a deep H$_2$-rich atmosphere, we demonstrate the capability of the scheme to reproduce non-condensing convection. We find convection occurring at pressures greater than 0.3 bar and the dynamical structure shows high-latitude prograde jets. Our convection scheme will aid in the 3D climate modelling of a wide range of exoplanet atmospheres, and enable further exploration of temperate sub-Neptune atmospheres.
\end{abstract}




\section{Introduction}

Over the past two decades, detection surveys have revealed thousands of exoplanets, most of which are between Earth and Neptune in size and in mass \citep{Fressin2013,Ricker2014,Fulton2018}. Having no solar system analogues, there are many open questions about them, notably what different atmospheric compositions and internal structures they may have. Different categories have been proposed: Super-Earths, Mini-Neptunes \citep{Valencia2007, Fortney2007, Seager2007, Valencia2013}, water-worlds \citep{Leger2004}, Hycean planets \citep{Madhusudhan2021}, and planets with magma oceans \citep{Schaefer2016}.
Knowing the exact makeup of individual planets has proved very difficult, as mass and radius measurements alone lead to degenerate internal structure solutions. Until recently there had been limited success in characterising the atmospheres of such planets. There have been suggestions of a `radius valley' separating super-Earths and mini-Neptunes, but also some statistical evidence for a third group of water-rich planets \citep{Luque2022}. 

The JWST, now in its third year of operation, promises to advance this frontier with its unparalleled precision and spectral range in the infrared. Transmission spectroscopy has been used to measure molecular abundances of $H_2$ dominated atmospheres on several sub-Neptunes \citep{Madhusudhan2023b, Benneke2024, Holmberg2024} and place constraints on the atmosphere of a possible ocean world \citep{Damiano2024}. With enough dedicated observing time and a better understanding of systematics, it might even be possible to characterise terrestrial exoplanets \citep{Meadows2023, Fauchez2022}. 

A particularly exciting prospect with JWST is its ability to study temperate sub-Neptunes. Whereas before only hotter planets with greater scale heights were accessible, now we can characterise atmospheres with Earth-like equilibrium temperatures. This opens up the possibility of, for the first time, being able to examine potentially habitable planets. The prime examples of this are candidate Hycean planets - a theorised category of sub-Neptunes with a thin hydrogen-rich atmosphere sitting atop a surface liquid water ocean which could be habitable under the right conditions. The hydrogen atmospheres mean that biosignature gases could be detectable even in ppm quantities \citep{Madhusudhan2021}.

To have a hope of effectively interpreting observations and answering these fundamental questions of habitability, a thorough understanding of planetary and atmospheric processes is required.
Observations with transmission spectroscopy, on their own, reveal information about the chemical abundances of various species around the 10-100 mbar level, about clouds and hazes, and give some rough indication of the temperature structure \citep{Madhusudhan2009, Madhusudhan2018, Benneke2013}. Beyond this is information we can only access indirectly: the detailed temperature structure, the atmospheric circulation patterns, deep cloud layers, the nature and presence of any surface, and ultimately the planetary habitability, which we restrict to meaning the presence of liquid surface water \citep{Barstow2020, Cockell2016, Meadows2018}. 
Understanding exoplanet atmospheric processes is what enables us to bridge this gap and go from a limited set of observables to detailed inferences about many aspects of the atmosphere, even if sometimes there are observational degeneracies which are hard to break \citep{Hu2021, Helling2019, Constantinou2024}. 
 
In the temperate sub-Neptune regime, particular care must be paid to correctly modelling processes which may behave in novel or unusual ways. One such process is atmospheric convection, which has a critical influence on an atmosphere's general climate state, and is expected to behave in very different ways across different sub-Neptune atmospheric regimes. Convection profoundly influences many different fundamental features of the atmosphere \citep{Pierrehumbert2010, Vallis2017, Iribarne1981, Emanuel1994, DelGenio2015}. It is one of the main controls on rainfall and clouds and has a controlling effect on the temperature structure. Due to the interplay between convective plumes and the general dynamical circulation, it also affects large scale dynamical and weather events. Convection is a complex and very highly non-linear process and can behave very differently in various planetary atmosphere regimes which are all expected to be found in various sub-Neptunes (e.g. \cite{Pierrehumbert2016, Ding2016, Leconte2017, Innes2023, Habib2024a, Leconte2024}). Understanding and modelling these variations is thus essential to properly predicting the nature of temperate sub-Neptune atmospheres.

On Earth, atmospheric convection, properly defined as the vertical mixing caused by density differences in an unstable medium \citep{Vallis2017}, generally starts when incoming solar radiation heats the surface and the air layers directly above it. This makes these layers unstable and gives rise to convective updrafts \citep{Emanuel1994}. As the plume rises, it entrains environment air with it. When (or if) it hits the Lifting Condensation Level (LCL), the plume becomes saturated and water condenses out into liquid or ice phase. The latent heat released encourages further convection.  Depending on the entrainment rate and the surrounding environment, convection can vary from shallow updrafts ending with minimal precipitation (cumulus clouds), to less-entraining plumes that can rise to the top of the troposphere as thunderstorm-causing cumulonimbus \citep{Lin2022}. 

Beyond this basic picture there is a great degree of complexity. Negatively buoyant downdrafts form, sometimes unsaturated but more commonly saturated. The broader environment subsides to maintain hydrostatic equilibrium. Downdrafts lead to cold pools - regions of lower potential temperature into which rainwater often evaporates - while updrafts seem to preferentially start on the edges of these pools. Convection can self-organise on larger scales, leading to meso-scale or even larger convective systems, all the way to supercell thunderstorms which provide a good fraction of the stratospheric water influx \citep{DelGenio2015, Arakawa2004, Rio2019}. 
Overall, convection is more efficient over land than water \citep{Takahashi2023}, and is strongly variable both temporally and geographically. It always transports heat and moisture upwards, and acts to keep the atmospheric profile relatively close to the moist adiabat.

This overall picture of moist convection rising from the surface may well be accurate for Earth-like planets, but it is not constant across different types of planetary atmospheres. A well-known different case is that of dry (non-condensing) convection. Here plumes form as in the moist case, but there is no condensation so no precipitation nor cold pools, although a deeper convective zone leads to more consistent self-organisation. Venus appears to have three different convective zones \citep{SanchezLavega2017} - there is sulphuric acid condensation in the highest zone but the latent heat release does not appreciably affect the dynamics of convection. For gaseous planets, a deep dry convection zone transports heat from the interior \citep{Heng2015, Fortney2021}, although for cool enough tropospheric temperatures, moist convection can happen there like with the gas and ice giants in our solar system. In the case of highly irradiated planets such as Hot Jupiters, conversely, radiative heat transfer is efficient and the top of the convective region can be hundreds of bars deep \citep{Gandhi2017}. Convection does not then have an impact on what is conventionally understood as the atmosphere.

A more complex case is what are known as "non-dilute atmospheres", where the mass mixing ratio $q$ of the condensible species before significant. The threshold has been set around $q=0.1$ (\cite{Pierrehumbert2016}, compare to a maximum $q$ on Earth of $\approx 0.02$). A common situation where this might happen is when a relatively thin background atmosphere, such as 1 bar of $N_2$ or CO$_2$, is in contact with a condensible surface reservoir, such as a warm liquid water ocean. Alternatively, we encounter these if the temperature is low enough for the main gas to be condensible, like on Mars \citep{Wordsworth2011}. Initially this leads to stronger and more variable convection, as described in \citet{Seeley2021}. High-resolution 3D modelling showed that for sea surface temperatures of $>320K$ in an Earth like atmosphere, convection enters a relaxation-oscillator like system with deluges every few days. For a 1 bar surface atmosphere, the strength of convection peaks for surface temperatures of $\sim 330K$ with $q\sim0.1$ \citep{Seeley2023}. But as $q$ rises after that, the atmospheric profile is constrained to be closer and closer to the saturation pressure-temperature profile $p_{sat}(T)$ and so convection becomes more sluggish \citep{Pierrehumbert2016, Ding2016}. It is harder to generate buoyancy and vertical motions are slower and more laminar. This also impacts on the global atmospheric circulation - the dynamics are more barotropic and the global temperature gradients are weaker.

Convection also behaves differently when the condensible species (for example water) is heavier than the background gas (e.g. an $H_2/He$ dominated atmosphere). Whereas in Earth's  atmosphere ($\mu_d = 28.96$) convection the higher-q parcel is made lighter by the presence of water ($\mu_v = 18.02)$, here the condensible can weigh it down and act against convection. If $q$ exceeds a critical value, convection is completely inhibited and another process - usually diffusion - transports heat and moisture up through the atmosphere \citep{Guillot1995, Leconte2017, Leconte2024, Habib2024b, Seeley2024}. This process is thought to create super-adiabatic layers in the Solar system gas giants as well as in some sub-neptunes \citep{Li2024, Leconte2024, Benneke2024, Clement2024}. Improperly modelling convection in these conditions can have immense effects on the temperature structure - off by as much as 150K at the 1 bar level in the case of \citet{Leconte2024}. It can also lower the runaway greenhouse limit and so push the inner edge of the habitable zone outwards for the affected planets \citep{Innes2023}.

Numerically modelling convection is most commonly done in the context of a General Circulation Model (GCM). GCMs are complex models which aim to simulate the whole atmosphere of a planet in 3D  \citep{Showman2013, Walters2019, Neale2010, Pierrehumbert2010}. They consist of, at a minimum, a dynamical core solving some form of the fluid equations of motion as well as some physics parametrisations. GCMs are widely used to model many aspects of exoplanet atmospheres such as their dynamical structure \citep{Innes2022, Hammond2020}, photochemistry \citep{Cooke2023},  clouds \citep{Charnay2021, Komacek2020}, and higher-level considerations of habitability such as demarcating the Inner Habitable Zone \citep{Kopparapu2016, Turbet2023, Kuzucan2025}.

In these GCMs, there are two main ways in which convection is usually represented: either with a mass-flux parametrisation or a convective adjustment scheme.
A convective adjustment scheme assumes radiative-convective equilbrium and relaxes an unstable thermodynamic profile to a marginally stable reference state (e.g. \cite{Manabe1965}). Conversely, a mass flux parametrisation simulates the statistical effect of convection plumes to obtain more accurate heating and drying rates across the convective region \citep{Arakawa2004}. There are several key aspects of these mass-flux schemes. The first is the guiding notion of quasi-equilibrium, that is that convection is loosely in balance with radiative and other large-scale forcing \citep{Arakawa1974}. The inclusion of a convective updraft is a necessity, as is corresponding large-scale subsidence, and most models also include negatively buoyant downdrafts. The inclusion of effects such as convective momentum transport allows convection to affect the large scale circulation in a more realistic manner. 

The parametrisation of entrainment and detrainment rates in a rising plume is another particularly important factor that has a major influence on the effects of convection. The weaker the entrainment rate, the higher the plume can rise as its excess buoyancy is not diluted by the entrained air. Conversely, stronger entrainment leads to shallow cumulus convection. Parametrisations often explicitly simulate these different varieties of convection, leading to multiple convection models within the same parametrisation.

Mass-flux schemes are much more accurate on Earth, where they reproduce detailed condensation and rainfall patterns \citep{Rio2019}. They are not perfect, however, and issues in representing convection and cloud feedbacks account for a significant fraction of the uncertainties in the IPCC global warming predictions \citep{Sherwood2014}. They are also tuned to represent Earth convection and so cannot necessarily be blindly trusted as the ground truth for exoplanet atmospheres \citep{Sergeev2024, Sergeev2022, Lin2022, DelGenio2015}.

Mass-flux schemes are the norm amongst Earth-focused GCMs \citep{Vallis2017, Rio2019}, and a mix of mass-flux and convective adjustment schemes are found for exoplanet GCMs \citep{Fauchez2020, Christie2022}. Some work has been done on investigating the differences this can lead to. \citet{Sergeev2020} looked at differences in convection schemes as applied to two terrestrial exoplanets, TRAPPIST-1e and Proxima b. They found this led to average surface temperature differences of $5-15$K alongside substantial differences in global temperature variations. In the Proxima-b case, the differences in convection scheme were enough to tip the atmosphere into different circulatory regimes. Either extratropical or equatorial Rossby waves were favoured, with corresponding differences in jet streams and equatorial super-rotation. This illustrates the significant effects of convection on planetary climates and the importance of as comprehensive convection modelling as possible.

Regardless of whether they are mass-flux or convective adjustment schemes, many GCM convection schemes are not capable of handling some of the more exotic modes of convection outlined earlier. Detailed work of this convection has thus mostly been restricted to high-resolution Convection Permitting Models (CPMs) which cannot cover the whole planet \citep{Leconte2024, Seeley2024, Habib2024b}, and so do not calculate the global climate self-consistently. To date, the LMD-G convective adjustment scheme has been adapted for convection in hydrogen-dominated atmospheres \citep{Leconte2024} and a convective adjustment scheme has been applied to a strongly non-dilute atmosphere using the Exo-FMS GCM \citep{Ding2016}. We believe that a GCM-suitable mass-flux convection scheme that can capture all of these behaviours would be a significant contribution to the field. It would allow to accurately simulate exoplanet atmospheres that are at the cutting edge of observations, and make inferences about their features with a greater degree of confidence. We set out to construct such a scheme. 

We use the open source ExoCAM \citep{Wolf2022} as our reference GCM. Specifically, we use the Zhang-Macfarlane scheme \citep{Zhang1995, Richter2008} as our starting point for the deep convection model, and the Hack scheme \citep{Hack1993} as the starting point for the shallow convection model. In Section \ref{sec:2} we outline some relevant thermodynamic basics that are commonly used in convection modelling. In Section \ref{sec:3} we present the details of the model and in Section \ref{sec:4} we see implementations on an aquaplanet TRAPPIST-1e case. In Section \ref{sec:5}, we apply the model to the sub-Neptune K2-18~b assuming a Mini-Neptune structure, before we summarise our results in Section \ref{sec:6}.

\section{Theoretical basics}
\label{sec:2}
In this section we review some of the fundamental theory behind convection. We first describe the basic theory around dry convection in Section \ref{sec:2_1}, then do the same for moist convection in Section \ref{sec:2_2}. In  Section \ref{sec:2_3}, we define and discuss varying quantities that are useful in parametrising convection.

\subsection{Dry convection in an atmosphere with compositional gradients}
\label{sec:2_1}

The basic requirement for convection in an atmosphere is that the density of an adiabatically lifted parcel falls faster than that of the surrounding air. For a non-condensing parcel in a medium with constant composition, as is very nearly true on Earth, this leads to the Schwarzschild criterion for convective instability

\begin{equation}
    \nabla_T > \nabla_{ad} \label{eq:Schwarzschild_criterion}
\end{equation}

where $\nabla_T = d \ln T / d \ln P$ is the atmospheric temperature gradient. The adiabatic gradient, assuming hydrostatic equilibrium, is

\begin{equation}
    \nabla_{ad} = \frac{\partial \ln T}{\partial \ln P}|_{ad} = \frac{R}{C_p} \label{eq:dry_lapse_rate}
\end{equation}

With $R=R_g/\mu$ the gas constant ($R_g$ is the universal gas constant and $\mu$ the mean molecular weight of the environment) and $C_p$ the specific heat capacity at constant pressure. Next, we can define the potential temperature as the temperature of a parcel if it were moved adiabatically to a reference pressure $p_0$. It is particularly useful as a convection diagnostic because $\Theta$ increasing (decreasing) with altitude means stability (instability) to convection:

\begin{equation}
    \Theta = T (\frac{p_0}{p})^{\frac{R}{C_p}} \label{eq:potential_temperature}
\end{equation}

When there is a compositional gradient in the atmosphere, the above equations are not sufficient. Assuming no mixing or significant precipitation in the parcel and so no parcel compositional changes, the requirement for convection becomes the Ledoux criterion:

\begin{equation}
    \nabla_T > \nabla_{ad} + \nabla_\mu \label{eq:Ledoux1}
\end{equation} 

With $\nabla_\mu$ the gradient in mean molecular weight. We will restrict ourselves to considering combinations of two gases, one of which we denote with the subscript $_d$ and the other with $_v$. By convention, $_d$ refers to the dry background gas and $_v$ to a possibly condensible secondary gas. To compute density differences between atmospheric elements, the standard approach is to use the virtual temperature $T_v$ defined such that $p = \rho R_d T_v$. Then

\begin{equation}
    T_v = T (1 - \bar{\omega} q) \label{eq:Virtual_Temperature}
\end{equation} 

where $q$ is the mass fraction of the second gas species and $\bar{\omega}$ is the reduced mean molar mass difference \citep{Leconte2017}. If $\bar{\omega}>0$ then the condensible species is heavier than the background gas, if $\bar{\omega}<0$ then it is lighter.

\begin{equation}
    \bar{\omega} = \frac{\mu_v - \mu_d}{\mu_v} \label{eq:mu_red}
\end{equation}

The Ledoux criterion can be written as

\begin{equation}
    \frac{d \ln T_v}{d \ln p} = \nabla_{T_v}> \frac{R_{m}}{C_{p,m}} \label{eq:Ledoux2}
\end{equation} 

$C_{p,m}$ is the composition adjusted heat capacity and $R_{m}$ is the composition adjusted gas constant. The $_{m}$ subscripts mean that these values are constructed from a mass-weighted average of the individual gas species values, $C_{p,m} = (1-q) C_p^d + q C_p^v$ and $R_{m} = R_g ( \frac{1-q}{\mu_d} + \frac{q}{\mu_v} )$.
The Ledoux criterion is very similar to the Schwarzschild criterion, with the only difference being the substitution of $T_v$ for $T$. Integrating Eq.\ref{eq:Ledoux2} leads us to a virtual potential temperature-like quantity $\Theta_v$:

\begin{equation}
    \Theta_v = T (1 - \bar{\omega} q )e^{- \int_{p_0}^{p} R_{m} / C_{p,m} \,d \ln p} \label{eq:virtual_potential_temperature}
\end{equation}

If $\Theta_v$ is constant in a layer, then this layer will be marginally stable to infinitesimal displacements. However, the changing environment $R_{m} / C_{p,m}$ means that neutral stability to infinitesimal displacements does not necessarily or usually imply stability to finite displacements, as discussed in \citet{Habib2024a}. This is an inherent feature of compositionally varying atmospheres: given that the Ledoux criterion as expressed in Eq \ref{eq:Ledoux2} incorporates the local $\beta_{m} = R_{m} / C_{p,m}$ as the determinant for stability, a rising parcel from elsewhere with a different $\beta_{m}$ will not be constrained by it. In practice, changes in $\beta_{m}$ are usually small and $\Theta_v$ is a good if not perfect descriptor of global stability in these cases.

\subsection{Moist convection}
\label{sec:2_2}

So far the composition of the parcel has been assumed constant as it is adiabatically lifted. However, when there is a condensing species in the parcel, this is not the case. The rising parcel may lead to some of the species condensing out of the gas and precipitating away from the parcel. 
Almost any species is potentially condensible, - whether or not it is actually is depends on the pressure-temperature regime it finds itself in. Common condensibles in planetary atmospheres include H$_2$O, CO$_2$, and NH$_3$. Without loss of generality, we will focus on H$_2$O as our default condensible species.

Commonly the compositional change in the rising parcel is neglected as the condensible fraction is low (the maximum $q$ found on Earth is very low, $\approx 0.02$ on a hot day in the tropics \cite{Pierrehumbert2010}) and water vapour and a $N_2$ or CO$_2$ dominated atmosphere do not have drastically different $\mu$. Despite this, a condensing species still substantially affects convection due to the latent heat released by condensation. This latent heat continuously warms the parcel, making it hotter and more buoyant as it goes up. This leads to the following moist adiabatic lapse rate for a saturated parcel \citep{Leconte2017}:

\begin{equation}
    \nabla_{ad}^* = \frac{R_{m}}{C_{p,m}} (1 + \frac{q_s}{1 - q_s} \frac{L}{R_d T})/(1 + \frac{q_s}{1 - q_s} \frac{L}{C_{p,m} T} \gamma_s) \label{eq:adiabatic_lapse_rate}
\end{equation}

which can be compared to the dry lapse rate in Equation \ref{eq:dry_lapse_rate}. $L$ is the latent heat of evaporation or sublimation. $\gamma_s = (\partial \ln q_s / \partial \ln T)_p = (1 - \bar{\omega} q_s) L / R_vT $. The subscript $_s$ denotes a quantity at saturation. This moist adiabat is shallower than the dry one, although it reduces to the latter as $q \rightarrow 0$, and it increases the range of atmospheric profiles that are unstable to convection. An atmosphere vulnerable to moist convection but not dry convection is said to be conditionally instable. An atmosphere with the vertical profile controlled by moist convection will have smaller temperature differences with height, compared to one controlled by dry convection.

Including the effects of compositional changes in the rising parcel can have important consequences. As detailed in \citet{Guillot1995} and \citet{Leconte2017}, the criterion for instability to moist convection which starts as 

\begin{equation}
    (\nabla_T - \nabla_\mu)_{env} > (\nabla_T - \nabla_\mu)_{parcel} \label{eq:Leconte17_14}
\end{equation}

by analogy to Eqs \ref{eq:Schwarzschild_criterion} and \ref{eq:Ledoux1} reduces to
\begin{equation}
    (\nabla_T - \nabla_{ad}^*)(q_s \bar{\omega} \frac{\mu_v L}{R_g T} -1) > 0 \label{eq:Leconte17_15}
\end{equation}

The multiplicative effect of the molecular weight changes is crucial. In the case where $\bar{\omega}>0$ (condensible is heavier than the background gas), this leads to moist convection being suppressed if $q$ exceeds a critical ratio:

\begin{equation}
    q_{crit} = \frac{R_g T}{\bar{\omega} \mu_v L} \label{eq:q_crit}
\end{equation}

\begin{figure}
    \centering
    \includegraphics[width=0.5\textwidth]{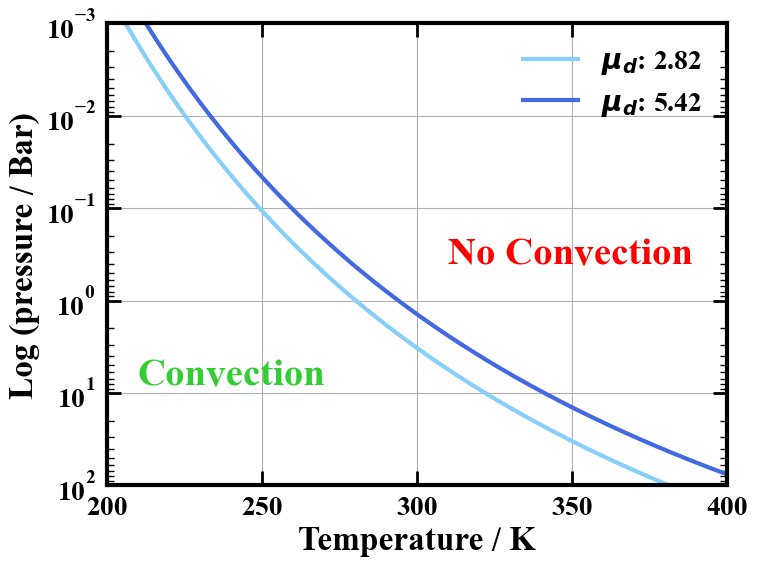}
    \caption{Moist convective inhibition in saturated atmospheres. Given a point in P-T space saturated with H$_2$O in a light background atmosphere, we calculate whether moist convection could occur. This is not the same as saying that moist convection would occur in these cases, which depends on a) the local atmosphere being at saturation and b) the local P-T profile being unstable. Note that dry (non-condensing) convection can happen across all of P-T space. The dividing lines are shown for two different dry atmospheric compositions. $\mu_d=2.82$ corresponds to a 10x solar metallicity atmosphere \citep{Charnay2021}, while $\mu_d=5.42$ corresponds to a 300x solar metallicity atmosphere quenched at 1000K and 100 bar \citep{Leconte2024}.}
    \label{fig:moist_convection_inhibition}
\end{figure}

We show the effect this has on moist convection in Figure \ref{fig:moist_convection_inhibition}, which shows the regions where moist (water) convection would be inhibited or not in a saturated atmosphere. The curves show the lines of marginal $q_{crit}$, where a saturated medium at that pressure and temperature will be neutral to convection irrespective of the local pressure-temperature profile. Regions with higher temperatures or lower pressures than this will have a higher $q_s$ and / or a lower $q_{crit}$ and so moist convection will be inhibited. The region with inhibited convection in a saturated medium is considerable: for a light enough background atmosphere, convection is inhibited at 300K and 1 bar: Earth-like conditions. 

It is important to note that this inhibition assumes that the P-T profile in the absence of convection (for example, the radiative profile) is steeper than the convective profile $(\nabla_T > \nabla_{ad}^*)$. In the reverse case - if the convective lapse rate is greater than the radiative one - the steeper of the two profiles is still favoured and we do in fact have convection.

\subsection{Useful quantities along a plume}
\label{sec:2_3}

When following a rising plume, two commonly used quantities are the dry static enthalpy $s$ and the moist static enthalpy $h_m$. The 1st law of thermodynamics and the associated Maxwell relations give \citep{Vallis2017} $d h = d \eta + \alpha dp$, where $h$ is the plume enthalpy, $\eta$ the entropy, $\alpha$ the specific volume and $p$ the pressure. Using the hydrostatic relation $\alpha dp = -g dz$ and the fact that $d\eta=0$ along the plume, we can write:

\begin{equation}
    d(h + gz)=0 \label{eq:dry_static_enthalpy}
\end{equation}

The enthalpy of a dry gas is $C_p T$, so we integrate, defining the \textit{dry static enthalpy} $s = C_p T + gz$ which is conserved in a dry rising plume.

In the condensing case, we define the latent heat of vapourisation $L = h_v - h_l$ where $h_v$ is the enthalpy of water vapour and $h_l$ the enthalpy of liquid water. $q_v$ is the water vapour mass fraction, $q_l$ the liquid water mass fraction and $q_t$ the total water mass fraction. Since $h_l = C_l T$ to a very good approximation, we can define the \textit{moist static enthalpy}.

\begin{equation}
    h_m = C_p^{dl}T + gz + Lq_v \label{eq:moist_static_enthalpy}
\end{equation}

with $C_p^{dl} = (1 - q_t)C_p^d + q_t C_l$. In the presence of a condensing substance, $h_m$ is conserved to first order in a rising parcel.

Dry and moist static enthalpies are precise enough to use as conserved quantities in cases when the condensible content of the parcel is low. However, when the water content $q_t$ becomes important, the precipitation losses mean that $h_m$ is no longer conserved and is less interesting as a variable to use in convection calculations. Instead we turn to the entropy $\eta$ as a variable to guide our ascending and descending plumes. Entropy - as we will see - is not conserved either in case of precipitation but the changes are at least easier to calculate.

The entropy $\eta$ of a substance is derived from its Gibbs Free Energy $g = -(\partial g / \partial T)_{p,q}$. In a dry mix of two non-condensing substances, it has the symmetrical form
\begin{equation} \begin{split}
    \eta = &(q_d C_p^d + q_v C_p^v) ln(T/T_0) - q_d R^d \ln(p_d/p_0) \\& - q_v R^v \ln(p_v/p_0) \label{eq:dry_mixed_entropy}
\end{split}\end{equation}

Where $T_0$ and $p_0$ are the reference values at which we set $\eta_0=0$.

When liquid and solid water are also present (even if only potentially as in the case of a parcel at, or near, the saturation level), the entropy of these components must be included. The entropy of the water vapour also changes in response to this potential liquid water. The general entropy expression can be written as

\begin{equation}
    \eta = (1-q^{vl})\eta_d + q^v \eta_v + q^l \eta_l \label{eq:general_entropy_expression}
\end{equation}

The entropy of liquid water is (up to a constant, and neglecting its small volume)
is 

\begin{equation}
    \eta_l = C^l \ln(T/T_0) \label{eq:liquid_entropy}
\end{equation}

To relate $\eta_v$ and $\eta_l$ we use a  similar method to that for the moist static enthalpy previously. Taking $L = T (\eta_v - \eta_l)$, which is true at equilibrium i.e. if the parcel is at saturation, we can write

\begin{equation}
    \eta_v = \eta_l + \frac{L}{T} \label{eq:relating_etav_etal}
\end{equation}

We include an extra $-q_v R_v \ln(e/e_s)$ term to account for the parcel not being saturated, following \citet{Iribarne1981}, and use the Buck equation for water saturation pressure \citep{Buck1981} as it gives the best results across the temperature range. 
Finally, for a fully general entropy expression we want to include the possiblity of solid condensate, such as ice, as well as liquid. Denoting the latent heat of vapourisation as $L^v$, we also introduce the latent heat of fusion as $L^f$. Accounting for this ice entropy, and denoting ice quantities by $^i$, leads to the following entropy expression which we use (very similar to \cite{Zeng2005}'s Equation 3.14):

\begin{equation}\begin{split}
    \eta_{dvli} =& ( (1-q^{vli}) C_p^d + q^{vli} C^l) \ln (T/T_0) - (1-q^{vli}) R^d \ln(p^d/p_0) \\& + \frac{q^v L^v}{T} - \frac{q^l L^f}{T}- q^v R^v \ln(e/e_s) \label{eq:entropy_final}
\end{split}\end{equation}

\section{An overview of the new scheme}
\label{sec:3}

The existing ExoCAM convection scheme is made up of two separate schemes, the Zhang-Macfarlane deep convection scheme (which we will hereafter refer to as the ZM scheme) and the Hack shallow convection scheme. To complete the moist physics parametrisation, these are combined with the cloud microphysics and macrophysics, which are responsible for computing the cloud cover and properties using a Sunqvist style scheme \citep{Sundqvist1988, Zhang2003}. Beyond that, the convection scheme couples with the rest of the sub-grid physics and the GCM's dynamical core. The atmosphere is split into a prescribed number of vertical layers. The default ExoCAM value for this is 40 \citep{Wolf2022} but this can be straightforwardly changed. The convection scheme is usually called every 30 model minutes along with the rest of the sub-grid physics parametrisation. The resulting forcing is split up and applied at successive intervals to the overall model physics state to match the dynamics, which run with a shorter time step \citep{Neale2010}.

The overall structure of the deep convection ZM scheme is fairly standard as far as mass-flux schemes go. Its main features are as follows:
\begin{itemize}
    \item Diagnose the convection start level.
    \item By following an entraining ascending plume, find the Convective Available Potential Energy (CAPE).
    \item Calculate entrainment rates and then find the detailed properties of updrafts and downdrafts in convective regions. This includes temperatures, precipitation, water content, and the relative mass fluxes in the plumes.
    \item From the updraft and downdraft quantities, find the heating and drying rates along the column.
    \item These heating and drying rates are then used to calculate the rate of CAPE destruction, which is used to fix the overall rate of convection and the final heating and drying rates.
\end{itemize}

The sections below describe the changes made to the Zhang-Macfarlane and Hack schemes in this current work. There are some elements of the model that are largely unchanged. These are the large scale heat and moisture transport equations, the transport of momentum and trace constituents, and the re-evaporation of precipitation (see \citet{Hack1993}, \citet{Zhang1995}, \citet{Richter2008} and \citet{Neale2010}).

Particularly in a non-dilute atmosphere \citep{Ding2016,Pierrehumbert2016}, significant precipitation can lead to several normally negligible effects becoming very important. Notably, the vertical movement of precipitation leads to vertical fluxes of mass and potential energy, and the reduction of the pressure above a particular layer causes it to expand, doing work. ExoCAM already handles the changes in mass distribution caused by precipitation. For the transport of potential energy, we assume that precipitation reaches its terminal velocity in a short time. This means that the potential energy liberated by precipitation falling through a layer is deposited in this same layer. This energy is the same magnitude as the pressure work done by the layer \citep{Ding2016}, and so the terms cancel out and no further parametrisation is required.

\subsection{Finding the deep convection base level}
A critical question is finding whether an atmospheric column is stable or unstable to convection, and in the unstable case finding the model layer where convection starts. This is commonly done using the moist static energy $h$: as this is conserved along an adiabatically rising parcel, $h$ decreasing with altitude indicates conditional instability. This method is flawed for two reasons. Firstly, because a layer with sufficiently high $q$ can be stable to moist but not dry convection, and secondly because $h$ is not strictly speaking conserved in a rising parcel with significant $q$.

We track the stability of the medium to both dry and moist convection separately. In the dry case - the condition is applied only in a subsaturated atmosphere - this is done using the virtual potential temperature $\Theta_v$ introduced in Eq. \ref{eq:virtual_potential_temperature}. We look for the lowest local maximum of $\Theta_v$.

\begin{equation}
    \Theta_v(p) = T(p) (1 - \bar{\omega} q(p)) e^{\int_{p_0}^{p} \frac{R_g}{\mu C_{p,mix}} \,d \ln (p)}
\end{equation}

A constant $\Theta_v$ with altitude does not denote the marginal stability of the system to finite but rather infinitesimal displacements (see \cite{Habib2024a}). This is the relevant condition for us as we are interested in the starting point of convection where the displacements would start.

In the moist case, we use the standard moist convective stability criterion in Eq. \ref{eq:Leconte17_15}. We again take the lowest layer unstable to convection as the convection start point. In cases with a surface liquid reservoir and $\bar{\omega}<0$ this will usually be the surface layer.

\begin{equation}
    (\nabla_T - \nabla_{ad}^*)(q_s \bar{\omega} \frac{\mu_v L}{R_g T} -1) > 0
\end{equation}

There is a degree of flexibility in when exactly to apply this moist convective condition. It needs to be applied when the system is somewhat subsaturated: the grid cells have significant internal variation in humidity and an overall unsaturated grid cell can have saturated regions. On Earth, almost all convection plumes start off dry i.e. subsaturated, and reach saturation at some point in their rise, which is what leads to the concept of an LCL in the first place. However, if the relative humidity criterion is too lax then we will see moist convection in unrealistically dry environments where condensation would not happen. We could set a minimum RH criterion for moist convection to trigger, but it is difficult to motivate a particular cutoff, as it ultimately depends on the amount of sub-grid variability in moisture as well as the amount of turbulence that decides the initial velocity distribution of small updrafts.
Instead, we use the existing model constraint that a plume must become buoyant in a few vertical layers. If it does not, and there is another suitable convection starting point as diagnosed by the equations above, then we follow another entraining plume from that point and may allow it to become the convection start point instead. This eliminates the chance that a unviable moist convective start point stops the presence of convection slightly higher up in the column.

We also allow a second convective start point and so a second convective region. It is a point of consideration whether there should be more than one convection start level. In the majority of cases, there will be only one convection layer - whether it starts at the surface or higher than that. In the event that some other layers are locally unstable, the shallow convection scheme can usually restore them to stability. However, in some cases, such as a temperate deep sub-neptune atmosphere (e.g. in \citet{Leconte2024}, a K2-18~b simulation with an  additional imposed albedo of 0.5), we find two convection layers (one lower region of dry convection, one upper region of moist convection) with a stable layer in between. In these circumstances it is desirable to have the deep convection scheme cover both these regions. If there exists an unstable layer above the first (lower convective region), then we treat that as a whole new starting point, calculating its plume evolution, CAPE destruction rate, convective mass fluxes and heating and drying tendencies independently from the lower convective region.

\subsection{Deep convection entrainment rates}
\label{sec:3_2}
An important part of the model is self-consistently calculating the overall entrainment rate $\lambda$ of a single model column. Given the horizontal extent of a single model cell (at a minimum, hundreds of kilometres) we need to represent a collection of plumes, each with its own entrainment rate. Some convective schemes, e.g. \cite{Gregory1990}, prescribe a single entraining rate for the cell, or vary it as a function of plume height, effective radius or mass. The Zhang-Macfarlane scheme takes a different approach and calculates the entrainment rates self-consistently.

For each level, there is a maximum entrainment rate $\lambda_D$ that can reach this level. A plume with a greater $\lambda$ would have entrained too much cooler material and would not reach this high. To find this $\lambda_D(z)$, \cite{Zhang1995} start off from the equation

\begin{equation}
    \frac{\partial h_u}{\partial z} = \lambda (h - h_u)
\end{equation}

with $h$ as the moist static enthalpy of the environment and $h_u$ as that of the updraft (from now on we drop the subscript $_m$, as we will not encounter standard thermodynamic enthalpy again). Using $h_u$ as the conserved variable in this way is troublesome because moist static energy is not conserved along the plume. We instead use a slightly modified entropy variable $\eta_t$ which is defined such that 

\begin{equation}
    \frac{\partial \eta_t}{\partial z} = \lambda (\eta - \eta_t)
\end{equation}

is true. We can follow a similar reversion of series procedure as ZM95 to arrive at the endpoint

\begin{equation}
    \lambda = \Delta + \frac{I_2}{I_1} \Delta^2 + \frac{(2I_{2}^{2} - I_1 I_3)}{I_{1}^{2}} \Delta^3 + ...
\end{equation}

where 

\begin{equation}
    I_n = \int_{z_b}^{z} ... (\eta_{t,b} - \eta_t)dz^{(n)}
\end{equation}

and 

\begin{equation}
    \Delta = \frac{\eta_{t,b} - \eta_{t,D}}{I_1}
\end{equation}

$\eta_{t,b}$ is simply the entropy at the plume base, as there has been no precipitation yet. $\eta{t,D}(z)$ is the $n_t$ of a plume element detraining at that height $z$. 

Previously the moist static energy of a detraining plume was set by assuming it had the same temperature (and, hence, density) as the surrounding air and was also saturated, so that $h_{m,D} = h_m^* = C_pT + gz + L q_{sat}$. The equivalence of temperature and density is no longer valid in compositionally varying atmospheres. We instead keep the requirement that the detraining plume element must be saturated and impose that the density of the detraining plume and environment must be equal, $T_D(1 - \bar{\omega}q_D) = T_e (1 - \bar{\omega} q_e)$. 

We also impose the requirement that the detraining $q_D$ cannot be greater than the plume base $q_b$ except in some rare circumstances near the cloud layer where we allow a slight excess. This allows us to find $s_D$, $h_D$ and $\eta_D$. We add an extra precipitation term of $(q_{p,b} - q_{p,sat})(C^l \ln(T_p/T_0) - \eta_D)$ to $\eta_D$ to give an adjusted entropy $\eta_t$ that includes the effect of precipitation on the specific entropy of the updraft. To treat some pathological cases near the plume top, we also require that the plume variables ($q_u$, $s_u$, $\eta_u$) cannot fall outside the maximum or minimum values encountered along the plume ascent, accounting for entropy changes as a result of precipitation). This should always be the case for an entraining-detraining-precipitating plume as long as the plume ascent is fast relative to the radiative timescale.

The Zhang-Macfarlane scheme also found it necessary to impose an upper limit on the entrainment of $2\times 10^{-4} m^{-1}$. The issue with this is that expressing it in $m^{-1}$ leads to very high entrainment values in a low $\mu$ atmosphere. We add a $\mu$ dependency to $\lambda_{max}$ so that the maximum entrainment rate in a layer stays constant across different atmospheric compositions.

\subsection{Tracking the deep convection updraft and downdraft properties}

\subsubsection{Updraft}
Following the Zhang-Macfarlane scheme, up to the Lifting Condensation Level (LCL, the level where condensation starts to happen) we track the evolution of plume quantities as follows:

\begin{equation}
    h_u^k = \frac{M_u^{k+1}}{M_u^k} h_u^{k+1} + \frac{dz^k}{M_u^k} (E_u h^k - D_u h_D^k) \label{eq:hu_updraft_calc}
\end{equation}

$k$ is the vertical index. The CAM vertical indexing works from the top down, so that $k+1$ is the level below $k$. $M_u^k$ is the updraft mass at $k$, normalised by the starting mass. $E_u$ and $D_u$ are the entrainment and detrainment rates in $m^{-1}$.

Once we reach the LCL, we can no longer use this due to precipitation losses. Instead we use a similar equation to find the new entropy $\eta_u^k$ pre-precipitation. Then, we invert the entropy equation using Brent's method \citep{Wolf2015} to find the temperature $T_u$. 

The fractional change in updraft parcel mass is given by 

\begin{equation}
    MS = \frac{1}{1 - R_u^k \frac{dz^k}{M_u^k}} \label{eq:mass_scaling_updraft}
\end{equation}

With $R_u^k$ the precipitation rate of the layer. $q_v$, $q_c$ and implicitly $q_d$ are modified by $MS$. The total mass of the plume is also modified by this amount - this introduces a new prefactor $M^*$ ($M^{*,k} = M^{*,k+1} MS^k$) that can be adjusted to represent successive precipitation losses. The precise calculation of $R_u$, $q_c$ and other quantities is covered in Appendix \ref{sec:a1}.

The important principle is that once plume water vapour has converted to liquid water, the precipitation of this water does not affect the plume properties (except for a minuscule change in pressure), and in particular does not change the temperature. We can then evaluate the entropy given the state variables $p, T, q_t$ to find the new parcel entropy. From these variables we also find the new $h_u$ and $s_u$. 
This is in contrast to the previous approach which used the assumption of $h_u$ being conserved and a saturated parcel to find the parcel properties. It is much more flexible across a range of conditions including significant $q_t$.

Previously the plume could end before its predicted end level if the plume $h_u$ went below the local saturated $h^*$. We implement a more appropriate ending condition - if the plume's virtual temperature $T_{p,v}$ falls below that of the environment $T_{e,v}$, such that the plume is no longer buoyant then the plume top is reached.

\subsubsection{Downdraft}
We make similar changes to find the downdraft properties. The downdraft is assumed to start at the level of minimum entropy, and to evaporate liquid water from the environment so that it remains saturated as it descends. This also helps keep it negatively buoyant. To find the correct temperature of the downdraft we run a similar method to inverting the entropy, solving for downdraft temperature $T_d$ such that

\begin{equation}
    \eta(p, T_d, q_s(T_d)) + \Delta q \eta_e^l = \eta_d \label{eq:ientropy_downdraft}
\end{equation}

$\Delta q$ is the additional water evaporated into the downdraft (normalised relative the downdraft mass post-evaporation) and $\eta_e^l = C^i \ln(T/T_0) + (q^l L^f / T)$ is the specific entropy of the environment condensed water. $\eta_d$ is the plume entropy calculated similarly to Eq. \ref{eq:hu_updraft_calc}. The size of the downdraft plume also increases as more water is evaporated in. The mass scaling factor becomes

\begin{equation}
    MS = \frac{1 - qd_{pre-e}^k}{1 - qd^k} \label{eq:mass_scaling_downdraft}
\end{equation}

with $qd^k$ the parcel's water vapour content after the evaporation and mass gain (when the parcel is saturated) and $qd_{pre-e}^k$ the parcel water vapour content before the evaporation and mass gain (subsaturated), 

Since downdrafts are usually triggered either by air motions next to cloud bases or by the evaporation of falling precipitation, we only allow downdrafts if there has been condensation in that column. This prevents downdrafts forming in dry convective cells.

\subsection{Changes to the Hack shallow convection scheme}

We also make changes to the Hack shallow convection scheme \citep{Hack1993, Neale2010}. The shallow scheme is used to represent the convective forcing associated with shallow and middle-level convection, commonly resulting in cumulus clouds, which is not treated by the deep convection scheme. The Hack scheme works somewhat similarly to a number of convective adjustment schemes by identifying pairs of unstable layers and moving them towards stability, along with some limited detrainment in a third layer above. This amount of detrainment, along with the total convective mass flux, are the main parameters that need to be determined.

The Hack scheme only focuses on plumes which rise only a small distance and so we are not overly concerned with potential changes in s and h along the plume. We change the initial detection of instability, comparing the virtual temperature of a layer with the virtual temperature of a parcel which has risen adiabatically (at constant $\eta$) from the layer below. We also allow for the existence of dry convection in the scheme which previously only allowed for condensing convection. The details of the dry convection implementation (a straightforward simplification of the moist convection case) are described in Appendix \ref{sec:a3}.

\section{Model validation: TRAPPIST-1e}
\label{sec:4}

We first test our model in a scenario where we are confident we know the desired results. In this regard, the exoplanet TRAPPIST-1e provides an excellent test case. Firstly, TRAPPIST-1e is an Earth-sized ($M_p = 0.772 M_\oplus$, $R_p=0.910 R_\oplus$) exoplanet with an instellation slightly lower than Earth ($\approx 900 Wm^{-2}$ vs $1366 Wm^{-2}$). As such, it is a planet where we might expect an Earth-like water cycle, with moist convection in a heavy background atmosphere and a low average $q$.

Additionally, another good reason to validate our model on TRAPPIST-1e is the existing benchmark from the THAI collaboration \citep{Fauchez2020,Turbet2022,Sergeev2022,Fauchez2022}. THAI was a GCM intercomparison program that compared the results of four GCMs (ExoCAM, LMD-G, ROCKE-3D, and the UM) in order to study the differences in how they modelled terrestrial exoplanets. This means that we have a wide range of results to compare to. In particular, we reproduce the Hab 1 case, whose key input parameters we show in Table \ref{tbl:hab1_params} \citep{Fauchez2020}.

\begin{table}
\centering
\begin{tabular}{cc}

\hline
\textbf{Atmosphere properties} & \\
Atmosphere dry composition & 1 bar $N_2$ + 400ppm CO$_2$ \\
Molecular dry air mass g mol$^{-1}$ & 28 \\
Atmosphere initial state & 300K isothermal; at rest \\
Atmosphere model top & 1 Pa \\
\hline
\textbf{Surface Properties} & \\
Surface composition & Slab ocean \\
Surface albedo (Liquid water) & 0.06 \\
Surface albedo (Ice or Snow) & 0.25 \\
Surface water layer depth & 100m \\

\hline
\end{tabular}
\caption[Atmosphere and surface parameters of the Hab 1 case]{Atmosphere and surface parameters of the Hab 1 case from \cite{Fauchez2020}}\label{tbl:hab1_params}
\end{table}

In this section we first look at the effect the new convection model has on the global scale temperature, wind and cloud patterns. We then look at how the convection operates on a more granular level. Overall, we show that convection behaves very similarly in our new scheme and the old one, as we would expect.

\subsection{Global climate}
\begin{figure}
    \centering
    \includegraphics[width=0.5\textwidth]{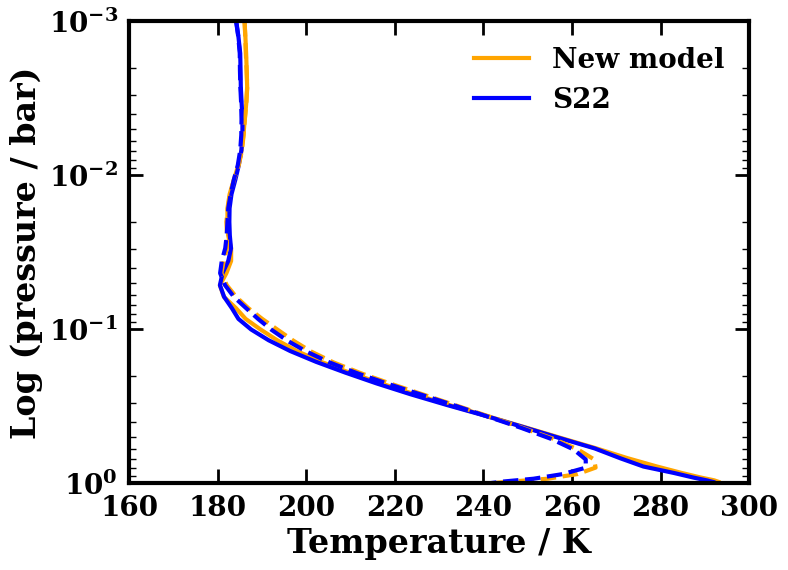}
    \caption{The TRAPPIST-1e Pressure-Temperature profiles at the substellar (straight lines) and antistellar (dashed lines) points, for our new scheme in orange and the ExoCAM results from S22 in blue. The PT profiles are similar in the stratosphere and upper troposphere before diverging near the surface. The lapse rate at the substellar point stays fairly constant, indicative of the effects of the moist convection. At the antistellar point, on the other hand, we see a pronounced temperature inversion. In both cases the difference between the new and the old convection models are minimal.}
    \label{fig:subanti_PTs}
\end{figure}

The global temperature structure of TRAPPIST-1e, as expected, presents significant day-night temperature differences both in previous work and with the new model. However, horizontal heat transport is efficient enough to maintain more temperate dayside temperatures, notably by preventing CO$_2$ freeze-out on the nightside. As shown in Figure \ref{fig:subanti_PTs}, the Pressure-Temperature profile of the atmosphere - one of the most important climate features - is very similar with the published results in \citet{Sergeev2022} (referred to as S22), with differences of $<2K$ that are straightforwardly attributable to time variability or simple stochasticity in the climate state. However, we do note a slight cooling in the upper troposphere that may be due to the inclusion of two convective layers, and a slightly warmer anti-stellar profile that is due to the increased flexibility in the convection start layer.

\begin{figure*}
    \centering
    \includegraphics[width=\textwidth]{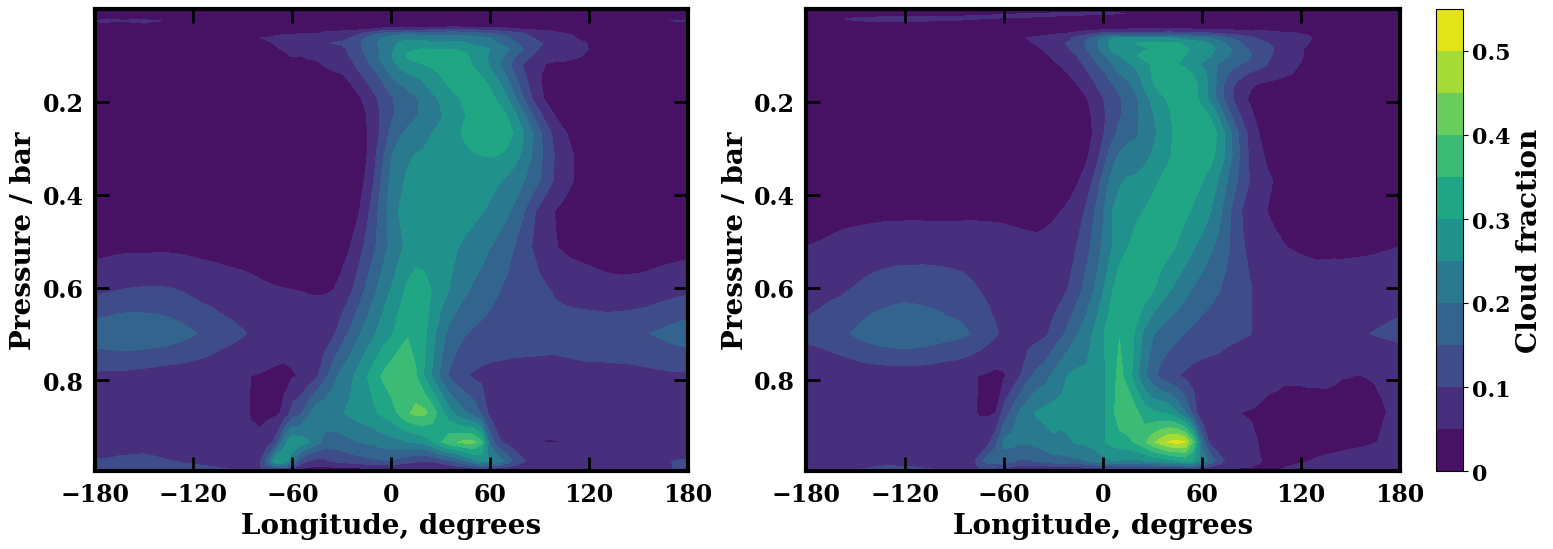}
    \caption{TRAPPIST-1e latitudinally averaged cloud fraction versus pressure. Left: Standard Zhang-Macfarlane and Hack schemes (Fig 7i, S22), Right: New scheme. The excellent similarities in the results can be easily noted. The bulk of the clouds are concentrated some way east of the sub-stellar point, with low and high altitude clouds being more eastwards than intermediate-altitude clouds. A characteristic anvil cloud shape can be seen at the cloud-top.}
    \label{fig:meri_cloud_compare}
\end{figure*}

We are also particularly interested in how the convection scheme affects the condensation patterns and cloud distribution. Clouds have a particularly important role on TRAPPIST-1e, as on other tidally locked exoplanets, as clouds are expected to concentrate on the dayside \citep{Sergeev2022}. This increases the total planetary albedo and drives the inner habitable zone inwards compared to what is expected in cloud-free models \citep{Kopparapu2016}. TRAPPIST-1e shows the expected pattern with dayside convection causing a thick cloud layer near the substellar point. Figure \ref{fig:meri_cloud_compare} shows a cross-section of the meridionally averaged cloud fraction, showing again the good performance of the new model in this test case.

\begin{figure}
    \centering
    \includegraphics[width=0.5\textwidth]{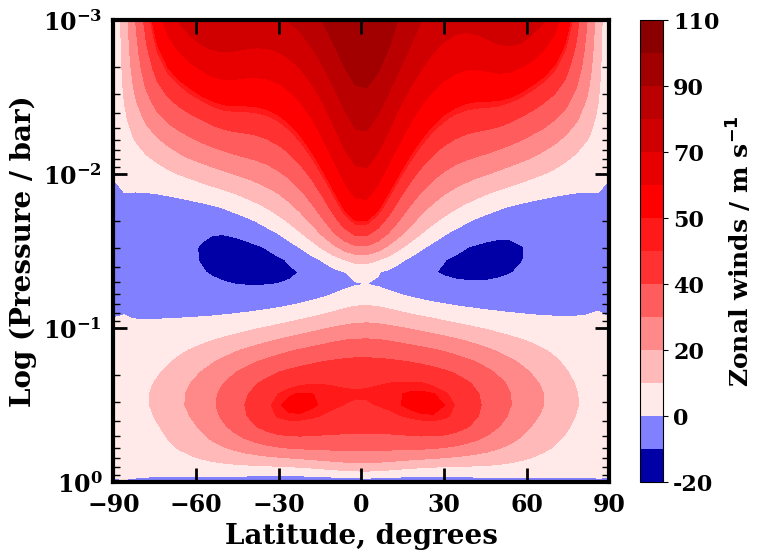}
    \caption{TRAPPIST-1e zonally averaged zonal winds with the new convection scheme. This plot shows a twin equatorial jet at low altitudes and another broad one at higher latitudes. These have a large impact on the global climate, contributing to heat and moisture transport to the nightside, climate time variability, and cloud patterns.}
    \label{fig:zonal_winds_new}
\end{figure}

As a check on the momentum transport of the convection scheme we show the mean zonal winds in Figure \ref{fig:zonal_winds_new}, which reproduce the expected shape and magnitude of  an equatorial jet structure with twin low-latitude jets \citep{Sergeev2022} as well as a single equatorial jet at high altitudes. We note that the twin-jets are marginally less apparent with our new model than in previous work. This is due to a difference of only a few m/s in the zonal winds and so may be due to simple randomness, although they may also be due to differences in convective momentum transport.

We impose both a 4th-order velocity damping and a 2nd-order velocity diffusion for the top layers following ExoCAM as described in \citet{Turbet2022}. We apply this damping to both our TRAPPIST-1e and K2-18~b test cases, but in the K2-18~b case we additionally change the value of the velocity diffusion coefficient \textit{del2coef} from $3.0 \times 10^5$ to $3.0 \times 10^4$ as part of the effort to contain angular momentum losses as discussed in Section \ref{sec:5_1}.

These figures all illustrate that the new convection scheme we have developed matches the state of the art existing convective parametrisations in the case of terrestrial exoplanet convection. This is an important step and confirms the utility of this model in a low-$q$, moist convective regime.

\subsection{Convection on TRAPPIST-1e}
\begin{figure}
    \centering
    \includegraphics[width=0.5\textwidth]{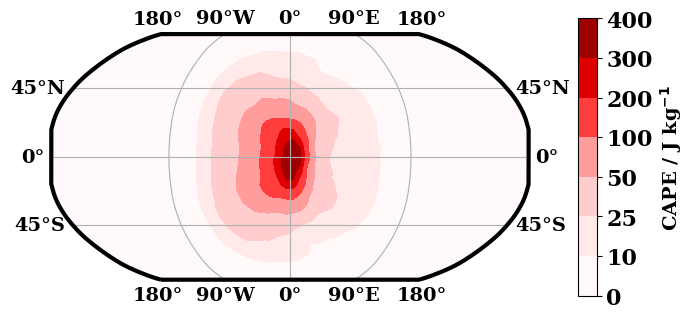}
    \caption{TRAPPIST-1e average CAPE values in $J kg^{-1}$. The average CAPE values increase sharply towards the substellar point, reflecting the sharp increase in the incoming solar radiation, particularly the amount deposited at the surface.}
    \label{fig:CAPE_new}
\end{figure}

Next we examine how convection behaves in detail in our simulations - where it happens, what layers of the atmosphere it affects, and what the overall heating and drying profiles are. Figure \ref{fig:CAPE_new} shows the average CAPE (Eq \ref{eq:CAPE}) value in a single model column. We see convection largely concentrated to the substellar region, with some more intermittent and weaker convection happening outside this region. This is as expected: for convection to develop we need atmospheric layers to become unstable, i.e. we need radiative heating. This exists only on the dayside and is strongest moving towards the substellar point, matching the CAPE trend seen.  The strength of convection is lower than Earth values, reflecting the weaker stellar forcing \citep{Zhang1995, Seeley2023}.

\begin{figure}
    \centering
    \includegraphics[width=0.5\textwidth]{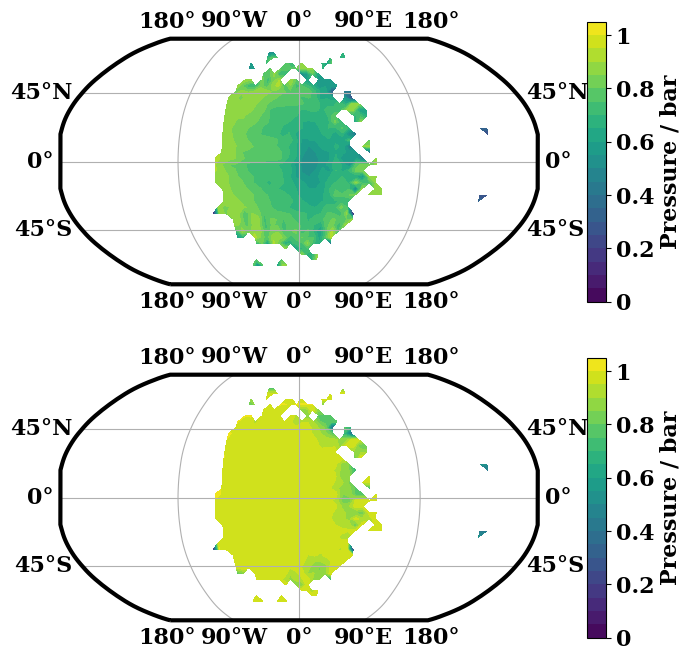}
    \caption{Upper and lower bounds of TRAPPIST-1e convection with the new scheme. Top: The average pressure where deep convection terminates. Bottom: The average pressure at which it starts. In the main substellar region, convection starts at the surface or very occasionally slightly above. Deep convection usually goes to the middle of the troposphere, similar to Earth. With stronger stellar forcing, deep convection goes to a higher altitude. Some isolated regions outside the substellar zone - some even on the nightside - see occasional convection. This convection is higher up, near the top of the troposphere.}
    \label{fig:pconvtb_avg}
\end{figure}

Figure \ref{fig:pconvtb_avg} shows the average pressures of the convection start and end levels respectively. As we would expect, convection usually starts at the surface where a large proportion of incoming stellar energy is deposited. The exception is the rare intermittent convection outside the substellar region, where occasional turbulence and dynamical effects can render regions higher up in the atmospheres convectively unstable. 

The (deep) convection top pressure varies more in time and in space. It sits at around 0.5 bar in the substellar region, equivalent to an altitude of roughly 5km. This is slightly lower than normal for deep convection on earth, and can be attributed to the lower maximum irradiation received \citep{Vallis2017}. Deep convection at the substellar point rises higher, a reflection of the greater heating there - the parcels near the surface receive the greatest amount of radiative heating as well as sensible heat from the surface, and so have the greatest buoyancy and can rise further before they hit the LNB (level of neutral buoyancy). We see only a few very rare and isolated instances of a second convective layer, and this upper convective region has a negligible impact on the local and global climate state. This is not a surprise given the expected presence of moist convection from the surface to the tropopause, which does not leave space for a second convective zone.

\begin{figure*}
    \centering
    \includegraphics[width=\textwidth]{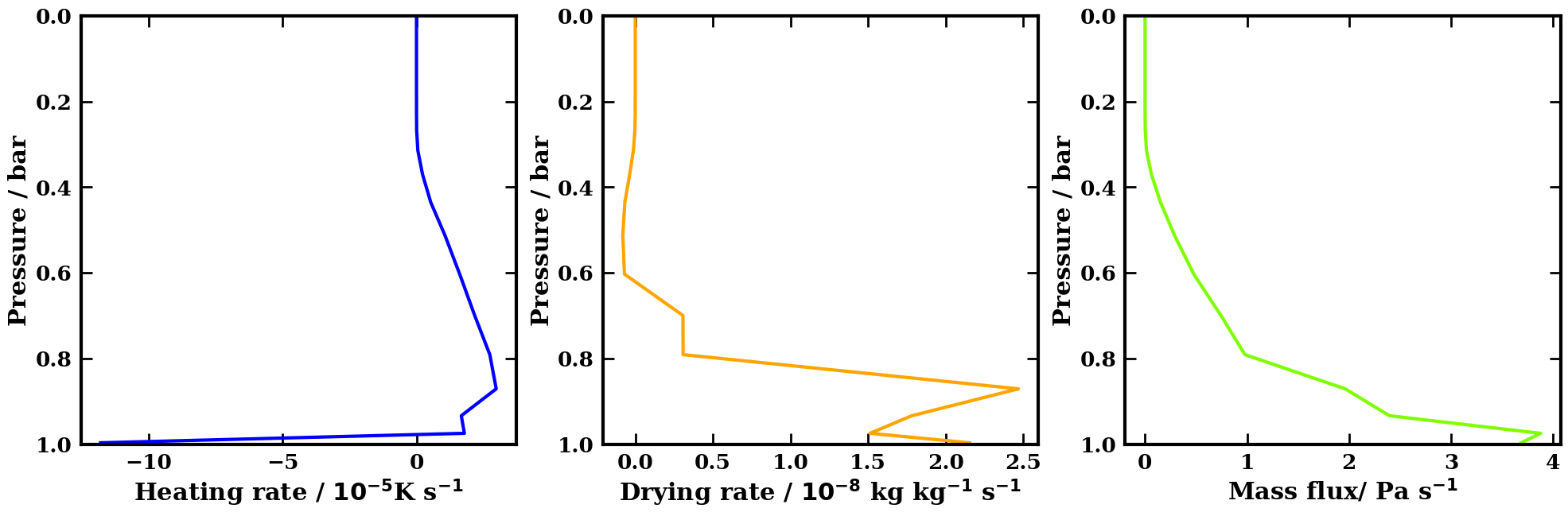}
    \caption{The effection of the new convection scheme on the TRAPPIST-1e climate state. Left: The average heating rate from deep convection when it is active. Centre: The average drying rate from active deep convection. Right: The average mass flux upwards (updraft - downdraft) from active deep convection. The heating rate is strongly negative at the model bottom where the updraft leaves from and quickly becomes positive, in large part due to the latent heat release from condensing water. Convection dries the lower and middle convective regions, but detraining saturated air moistens the upper convective regions. A net heating and drying of the overall column can be observed: this is due to the condensation and precipitation of water out of the atmosphere.}
    \label{fig:zmdtdq_new}
\end{figure*}

To examine the transport of moisture and heat, we average the heating and drying rates across the substellar region, as shown in Figure \ref{fig:zmdtdq_new}. As expected, convection transports both heat and moisture upwards. The profiles are similar to those on Earth - drying and cooling at the bottom of the column, with the heat and moisture deposited towards the top. Reflecting the fact that heating (from direct radiation and contact with the surface) is concentrated in the surface layer, the cooling for this layer is much more pronounced than in the rest of the atmospheric column. Convection itself leads to an overall drying and heating of the water column. This is expected as the rain water total, and the corresponding energy required to evaporate it, are accounted for separately.

\section{Model validation: Mini-Neptune Scenario}
\label{sec:5}

We would also like to see how the convection scheme performs in a different region of parameter space, specifically in somewhat warmer atmospheres with a higher condensible content and higher pressures. We choose the bulk parameters of the temperate sub-Neptune K2-18~b \citep{Foreman2015, Montet2015} for this test case. K2-18~b is a sub-Neptune ($R_p = 2.61 R_\oplus$, $M_p = 8.63 \pm 1.5 M_\oplus$) orbiting a M-IV Dwarf and receiving almost exactly the same instellation as the Earth ($1370$ W m$^{-2}$), T$_{\rm eq}$=250 K for a Bond albedo $A_b=0.3$ \citep{Cloutier2019}.
K2-18~b is the archetypical Hycean candidate \citep{Madhusudhan2021}, a proposed class of habitable sub-Neptune planets with deep water oceans under a $H_2$ dominated atmospheres. 

K2-18~b has been the subject of recent JWST observations \citep{Madhusudhan2023b}. The presence of CO$_2$ (to $5 \sigma$) and CH$_4$ ($3 \sigma$) but non-detection of NH$_3$ is consistent with the presence of an ocean surface at relatively low pressures \citep{Hu2021, Madhusudhan2023a}. Given the mass-radius constraints on the internal structure planet, such a surface could be made of liquid water \citep{Madhusudhan2020, Rigby2024}. However, other theories have been put forward to explain the atmospheric chemistry, such as reduced vertical mixing caused by moist convective inhibition \citep{Leconte2024}. In this case, moist convective inhibition leads to the formation of a stable layer with a much greater temperature and compositional gradient, and resulting higher temperatures at higher pressures. A liquid water ocean would be possible with a 1 bar surface pressure if $A_b \geq 0.55$.

For the purpose of this work, we assume a planet with the bulk parameters of K2-18~b and with a very deep atmosphere, as in \citet{Charnay2021} and \citet{Innes2022}. This allows us to study the behaviour of deep dry convection in an atmosphere with a significant condensible fraction. Overall, the global climate is reasonably similar to that found in \citet{Charnay2021} and the circulatory regime similar to \citet{Innes2022}. We divide our results into three parts: the global temperature structure, the global circulation regime, and then the convection and clouds structure.

\subsection{Details of model setup}
\label{sec:5_1}
We run a similar model setup to \citet{Charnay2021}. We assume a 100x solar metallicity atmospheric composition with the molecular abundances adopted from the \cite{Christie2022} Case 3 values, and use the planet parameters as shown in Table \ref{tbl:c21_params}. We assume it is tidally locked, which is probable, and so its orbital period of $33$ days is also its rotation period. We run the ExoCAM GCM with our new convection scheme in a 51-layer setup, simulating only the top 10 bars of the atmosphere for reasons of computational efficiency and cost. We set the model bottom layer $q$ to 0.22 following \citet{Charnay2021}. This enables us to probe deeper and hotter conditions with higher $q$. The key simulation parameters are summarised in Table \ref{tbl:c21_params}. 

\begin{table}
\centering
\begin{tabular}{cc}

\hline
\textbf{Planet Properties} & \\
Radius (m) & $1.66 \times 10^7$ \\
Rotation period (days) & 32.92 \\
Gravity (m s$^{-2}$) & 12.4 \\
Instellation (W m$^{-2}$) & 1370 \\
\hline
\textbf{Atmosphere properties} & \\
Molecular dry air mass (g mol$^{-1}$) & 4.01 \\
Molecular dry $C_p$ ($J kg^{-1}K^{-1}$) & 7952 \\
Atmosphere initial state & ExoRT 1D profile; at rest \\
Bottom layer $q$ ($kg kg^{-1}$) & 0.2 \\
Internal Temperature $T_{int}$ ($K$) & 30\\
'Surface' layer heat capacity ($J K^{-1}$) & $4.1\times 10^{6}$ \\
\hline
\textbf{Simulation properties} & \\
Model top pressure & 10 Pa \\
Model bottom dry pressure & $10^6$ Pa \\
Atmosphere layers & 51 \\
Runtime (Earth days) & 40,000 \\
\hline
\end{tabular}
\caption[Simulation parameters of the our mini-Neptune K2-18~b run]{Simulation parameters of our mini-Neptune K2-18~b run inspired by \cite{Charnay2021}}\label{tbl:c21_params}
\end{table}

Initially, the model runs encountered non-conservation of angular momentum leading to counter-rotating jets similar to those in \cite{Christie2024}. AM conservation in CESM's finite volume dynamical core is a known issue \citep{Lauritzen2014, Lebonnois2012}. It is especially problematic in configurations with slow rotators and with reduced surface torques, both of which apply to us as our surface has no height variations. Interesting, the AM issues only seem to become significant when portions of the atmosphere exceed $\approx 600K$. Why this is the case is not clear, but it may be linked to different atmospheric dynamics emerging in these hotter cases.

To address this issue, we implement the recommended changes from \citet{Toniazzo2020}. They located the main source of AM non-conservation, and developed a correction for this as well as a fixer to conserve global angular momentum in the dynamical core. This gives us much improved AM conservation in our model runs.

\subsection{Thermal structure}
\begin{figure}
    \centering
    \includegraphics[width=0.5\textwidth]{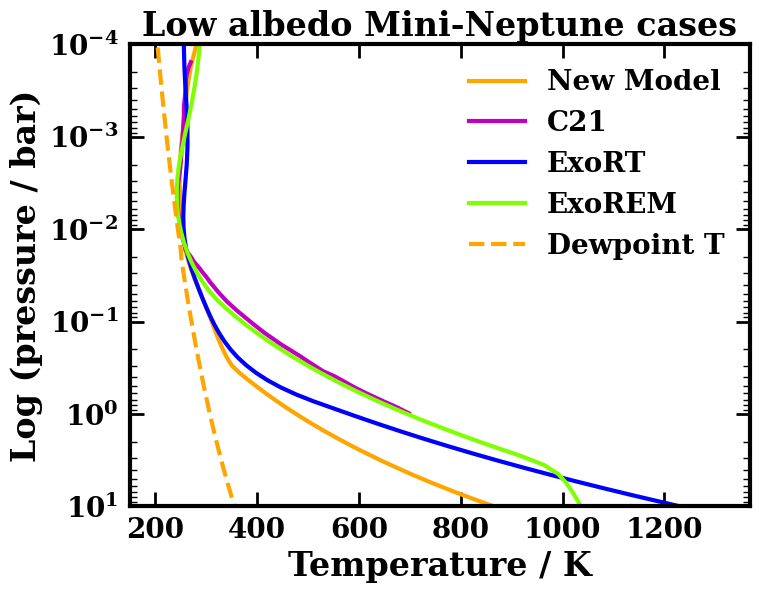}
    \caption{Pressure-Temperature profiles of Mini-Neptune K2-18~b cases: hypothetical scenarios with a deep atmosphere and no imposed albedo or hazes. New model: global average of our GCM run. C21: \citet{Charnay2021} 100x metallicity, tidally-locked case. ExoRT: 1D Radiative model of the 100x metallicity atmosphere. Note that this is in radiative equilibrium, not radiative-convective equilibrium, and so the temperature profile of the lower atmosphere is not expected to be accurate. ExoREM: 1D radiative-convective model, 100x solar metallicity case from \citet{Charnay2021}. Dewpoint T: the dew-point temperature and how it varies with pressure (and the water content at that pressure). In our run we can clearly see the convective zone below 0.3 bar.  The climate state is too hot by a few Kelvin to allow for clouds except in a few localised cold-spots, in contrast to the \citet{Charnay2021} results, although this is not apparent in this plot due to the thickness of the lines. There are very few geographical temperature variations, confirming that the atmosphere exists in a Weak Temperature Gradient regime. The ExoRT and ExoREM P-T profiles show that the difference in the GCM P-T profiles are mainly a result of the different RT schemes used.}
    \label{fig:k2-18b_selected_PTs}
\end{figure}

Figure \ref{fig:k2-18b_selected_PTs} shows the Pressure-Temperature profiles at the antistellar and substellar points. We also plot the P-T profile from the \citet{Charnay2021} 100x metallicity case. Overall, our P-T profile is about the same temperature at lower pressures, and significantly cooler at higher pressures. Our profile does not show the drop in lapse rate at high pressures that the C21 profile shows (when that profile becomes near isothermal). This is due to the different radiative transfer codes used: a 1D version of ExoRT we ran also had this continued high lapse rate. To illustrate this, we plot this 1D ExoRT \citep{Wolf2022} version as well the 1D radiative-convective model ExoREM which uses the same RT scheme as \citet{Charnay2021}. We note that the ExoRT profile plotted is radiative only, which explains it going to unrealistically high temperatures at high pressures. Its purpose is to show that the differences in GCM P-T profiles are driven by differences in the radiative transfer schemes. The difference in the performance of the two radiative schemes in the upper troposphere is driven largely by the different opacity sources considered. Specifically, the additional inclusion by ExoREM of NH$_3$, CO, and PH$_3$ gas opacities make a noticeable difference, as does the inclusion of H$_2$-He CIA.

Apart from some differences in the upper atmospheres - a temperature inversion caused by absorption in the NIR is present on the dayside but not the nightside - there are few horizontal variations across the whole planet. This is unsurprising: the previous GCM simulations of sub-Neptunes, especially temperate ones, also find minimal temperature differences, especially in slow rotators like K2-18~b \citep{Christie2022}. 
Theoretically, this can be explained by placing K2-18~b in the "Weak Temperature Gradient" (WTG, \cite{Pierrehumbert2016,Pierrehumbert2019}) regime, where a small Coriolis force leads to an necessary reduction in horizontal pressure and so temperature gradients. This regime is present to an extent in the tropics on earth where the Coriolis force is always small, but is global on a slow-rotator like K2-18~b. We note that being in the WTG regime cannot be fully predicted for a given planet from its basic parameters as the average zonal speed \textit{U} is an important parameter, but the emerging body of GCM simulations suggest K2-18~b very much is in this regime.

\subsection{Global Circulation regime}
\begin{figure}
    \centering
    \includegraphics[width=0.5\textwidth]{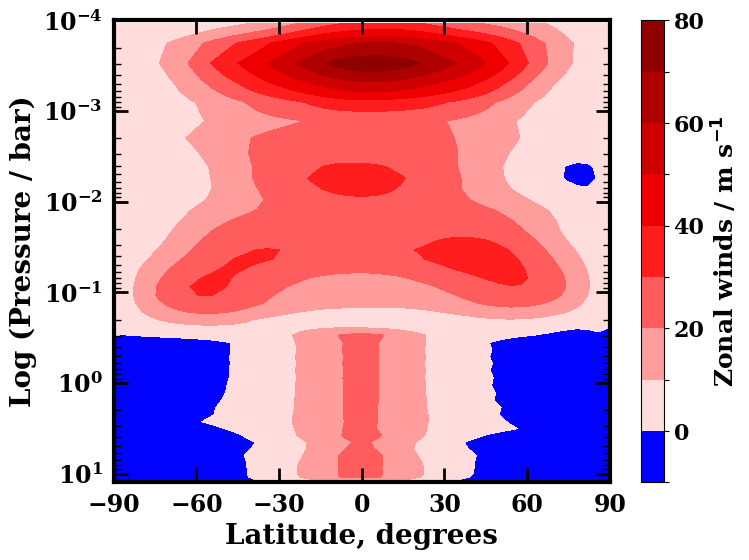}
    \caption{K2-18~b zonally averaged zonal winds, showing twin super-rotating jets at intermediate altitudes becoming a single equatorial jet at high altitudes, and a weak super-rotating equatorial tropospheric jet.  The near-surface average velocities are low due to the drag imposed at the model bottom.}
    \label{fig:k2-18b_zonalwinds}
\end{figure}

The zonal-averaged zonal winds are shown in Figure \ref{fig:k2-18b_zonalwinds}, showing an equatorial super-rotating jet in the upper troposphere, transitioning into two super-rotating jets at mid-latitudes around the 0.1 bar level. 
In the lower atmosphere, we see a single narrower equatorial jet with some retrograde motion at higher latitudes. The overall atmosphere has positive angular momentum, which is plausible given the presence of surface drag, which can add net angular momentum to the atmosphere.

\begin{figure*}
    \centering
    \includegraphics[width=\textwidth]{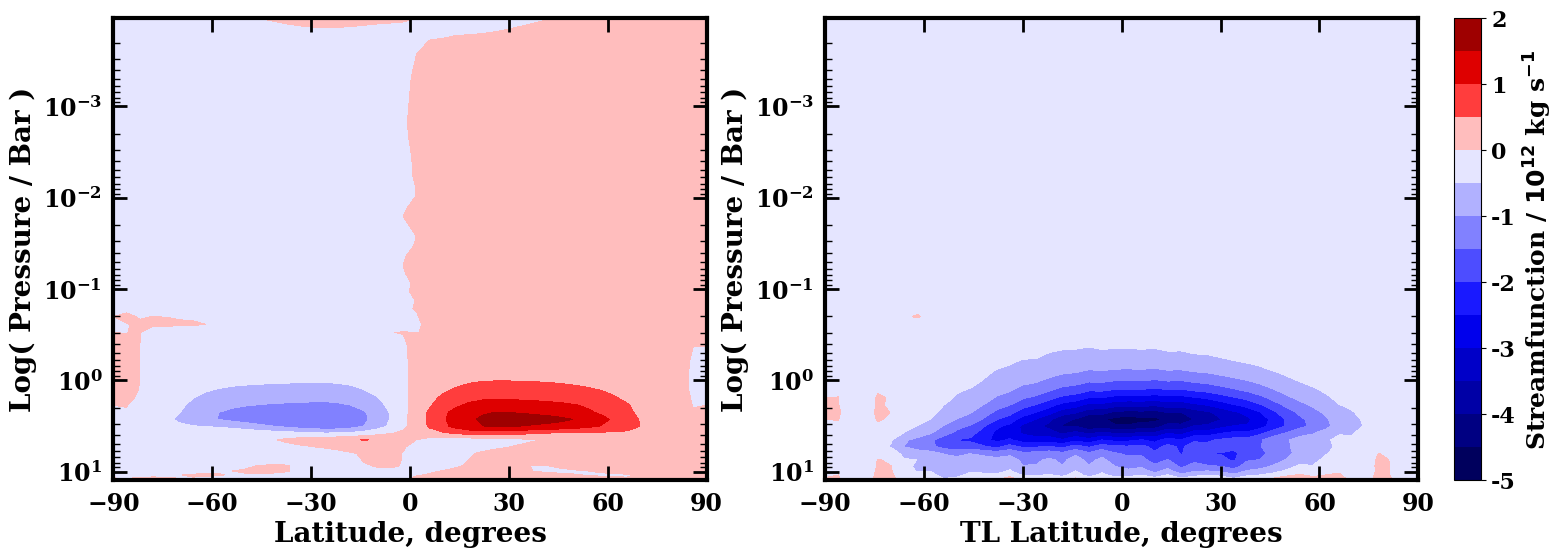}
    \caption{Left: K2-18~b mean meridional mass streamfunction in the standard latitude-longitude coordinates, in units of $kg$ $s^{-1}$. Right: K2-18~b mean meridional mass streamfunction in Tidally-locked latitude-longitude coordinates, in units of $kg$ $s^{-1}$. A TL latitude of 90 (-90) $\degree$ corresponds to the substellar (antistellar) point. Average circulation is anticlockwise around negative values and clockwise around positive values. The slow rotation speed of K2-18~b means that we see a single Hadley dominant cell going from the equator to the poles. We see that the dayside-nightside overturning circulation is stronger than the pole-equator overturning circulation, but of the same order of magnitude.}
    \label{fig:k2-18b_msf}
\end{figure*}

\begin{figure}
    \centering
    \includegraphics[width=0.5\textwidth]{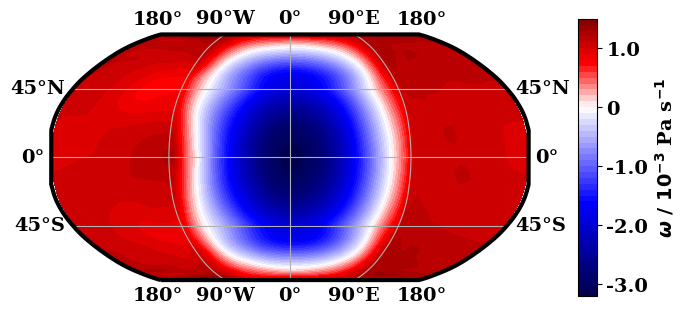}
    \caption{K2-18~b vertical velocity $\omega$ at the 0.01 bar level. Negative (positive) $\omega$ indicates upwards (downwards) motion. The global overturning circulation is clearly visible. Irradiation concentrated on the dayside heats and expands these atmospheric regions, driving upwards motion. There is corresponding subsidence on the nightside.}
    \label{fig:k2-18b_omega}
\end{figure}

Studies of tidally-locked (TL) exoplanets \citep{Hammond2020} show that the TL overturning circulation can be the dominant global circulation, which also affects heat transport. 
To compare overturning circulations, we look at the mean meridional overturning circulation as defined in \citet{Innes2022}.
We look at the mass streamfunction in both the normal latitude-longitude coordinates  and also the TL latitude-longitude coordinates (TL North Pole - substellar point, TL South Pole - antistellar point) in Figure \ref{fig:k2-18b_msf}.We see both the pole equator overturning circulation and the stronger tidally locked overturning circulation, which is responsible for the net dayside-nightside energy transport. To support our impression of a dayside-nightside overturning circulation, Figure \ref{fig:k2-18b_omega} shows the vertical velocity at 0.01 bar. Most of the dayside region is rising, with compensating subsidence on the nightside.

\begin{figure*}
    \centering
    \includegraphics[width=\textwidth]{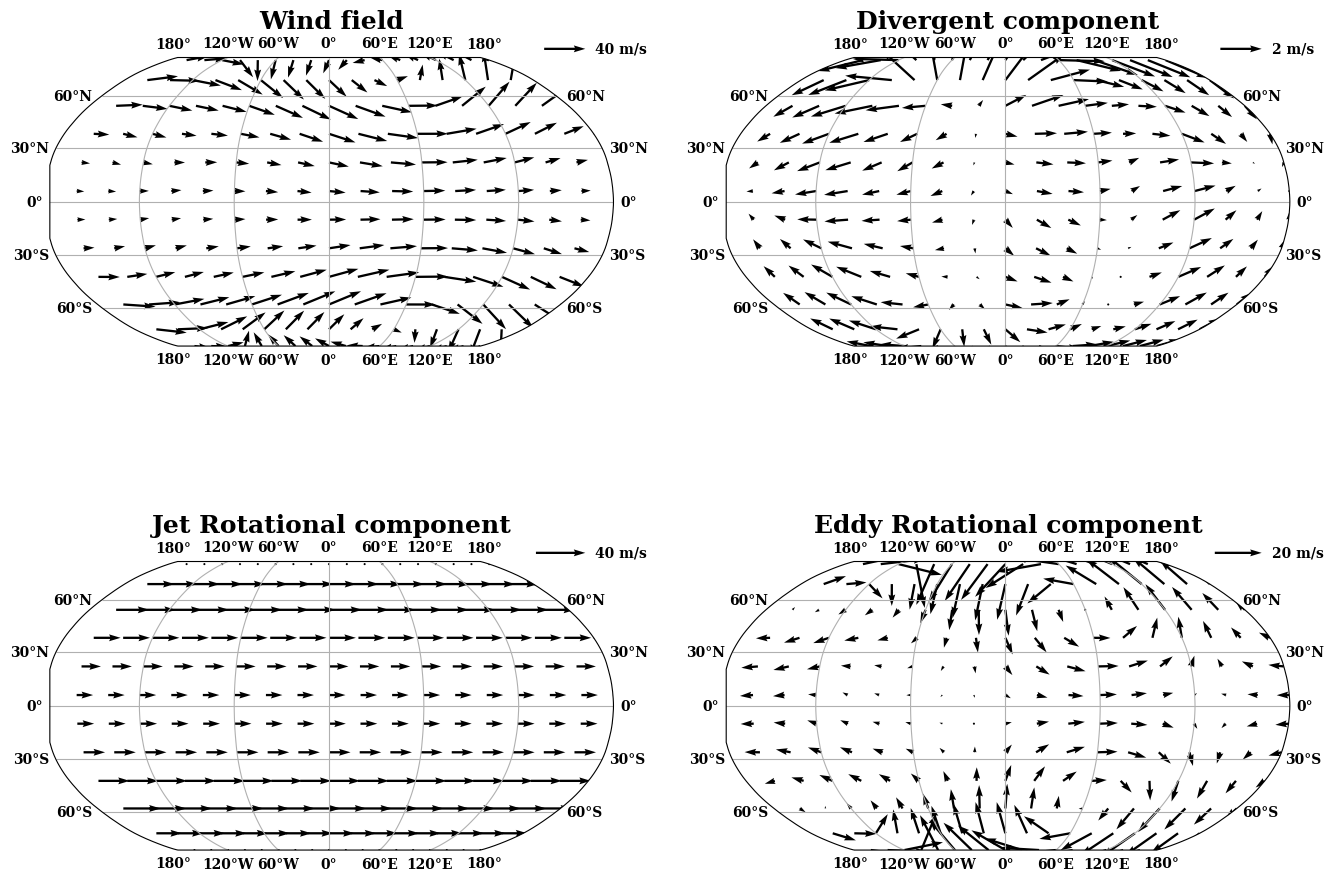}
    \caption{K2-18~b: Helmholtz decomposition of the wind field. Top left: horizontal wind field at 0.1 bar. Top right: Divergent component of the circulation. Bottom left: Zonal-mean rotational (Jet) component Bottom right: Eddy rotational component. The eastward nature of the overall winds can be seen, although it is strongest moving from the substellar  to the antistellar point. The wind's jet rotational component makes up the majority of the wind and shows twin high-latitude jets. The eddy rotational component is also significant and shows planetary scale standing waves of order 1, strongest in the eastern hemisphere. Compared to the rotational component, the divergent circulation is small and shows broadly isotropic substellar $\rightarrow$ antistellar point motion, although this is compensated by a returning circulation at lower pressures. Despite its small magnitude, the divergent circulation is still responsible for the majority of the net dayside-nightside heat transport.}
    \label{fig:k2-18b_helmholtz}
\end{figure*}

Another way of examining the dynamical structure is to perform a Helmholtz decomposition of the wind into its divergent, zonal-mean rotational (i.e. zonal jet), and eddy rotational components, as show in in Figure \ref{fig:k2-18b_helmholtz} which shows these components at the 0.1 bar level. Firstly, we see a roughly isotropic dayside-nightside divergent circulation which matches our expectations from the overturning circulation seen in Figures \ref{fig:k2-18b_msf} and \ref{fig:k2-18b_omega}, but with a magnitude much lower than that of the rotational circulations. The eddy-rotational component, which corresponds to standing waves, is important and clearly shows the presence of planetary-scale stationary waves of order 1.

Even if the divergent circulation is small compared to the rotational one, it can still have a leading role in transporting heat. To assess this, we follow the same technique as in \citet{Hammond2021} in calculating longitude-averaged heating and cooling from radiation and recirculation. Column integrated, the local energy budget is:

\begin{equation}
    <\nabla \cdot s\underline{u}> + F_S - F_{OLR} = 0
    \label{eq:su_div}
\end{equation}

\begin{figure}
    \centering
    \includegraphics[width=0.5\textwidth]{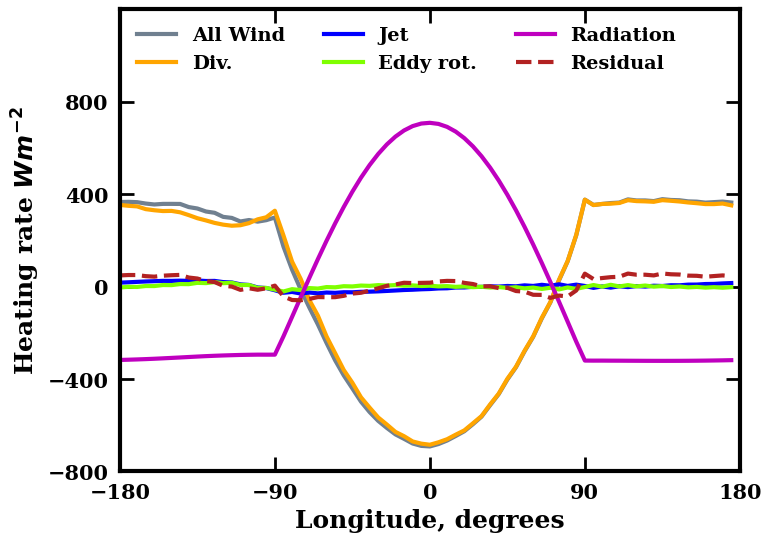}
    \caption{K2-18~b: The contributions of stellar heating and atmospheric circulation to the large-scale heating patterns, as shown for each longitude slice, averaging over all altitudes and latitudes. The divergent circulation clearly makes up the vast majority of the day-night heat flux despite its much smaller magnitude. There are some residuals left from regridding approximations.}
    \label{fig:k2-18b_divsu}
\end{figure}

The various meridionally-averaged components of this equation are shown in Figure \ref{fig:k2-18b_divsu}. The divergent circulation overwhelmingly dominates the day-night heat transport, and in fact the jet rotational and eddy rotational components are negligible compared to the divergent circulation. As discussed in \citet{Hammond2021}, this dominance can be explained by the planet being in a WTG regime. Expanding $\nabla \cdot s\underline{u} = s \nabla \cdot \underline{u} + \underline{u} \cdot \nabla s$, the irrotational component of the wind will have no divergence, eliminating the first component, whilst the second component is small due to the WTG regime. In contrast, the divergent portion of the wind will have a significant $s \nabla \cdot \underline{u}$ term and so will provide the majority of the heat transport.

\subsection{Convection}
\begin{figure}
    \centering
    \includegraphics[width=0.5\textwidth]{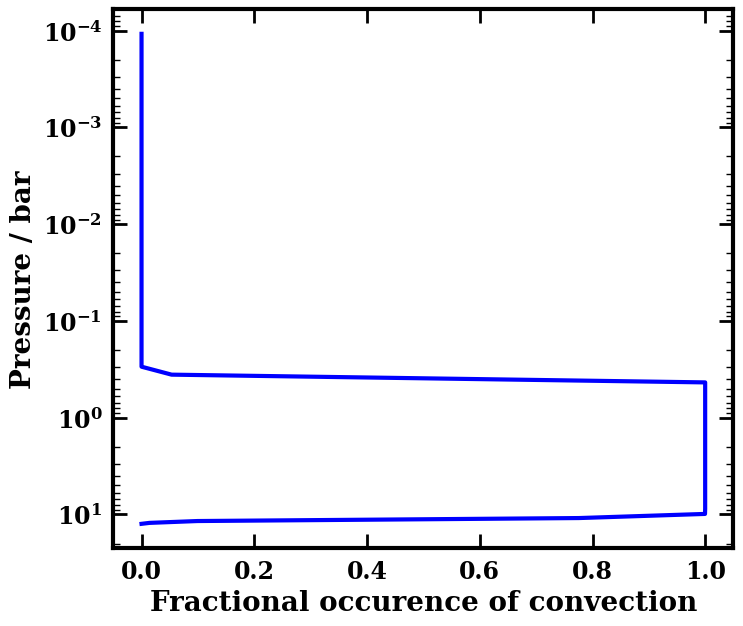}
    \caption{K2-18~b: across the pressure levels simulated, the fraction of the time that deep convection is active. We see universal convection between 10 bar (the model bottom) and 0.3 bar, with some intermittent convection around at these upper and lower boundaries: not every convective plume starts and ends at exactly the same level. This universality of convection is expected given the minimal geographic variation in pressure-temperature profiles: between 0.3 and 10 bar, the radiative lapse rate exceeds the convective one everywhere, and so dry convection sets in.}
    \label{fig:k2-18b_columnconv}
\end{figure}

To dive into convection, we first look at the fractional occurrence of convection happening at different pressure levels, as shown in Figure \ref{fig:k2-18b_columnconv}. We see universal convection happening throughout the atmosphere between the 10 bar and 0.3 bar levels. There is some intermittent convection happening around these top and bottom convective boundaries: not every convective plume starts or ends in the exact same place. We reproduce qualitatively the radiative-convective-radiative structure commonly seen in temperate sub-Neptune atmosphere modelling, e.g. the ExoREM P-T profile in Figure \ref{fig:k2-18b_selected_PTs}. We do not seen much of this lower radiative zone because the bottom of our convective zone is just above our model bottom, but we can establish its existence.

\begin{figure}
    \centering
    \includegraphics[width=0.5\textwidth]{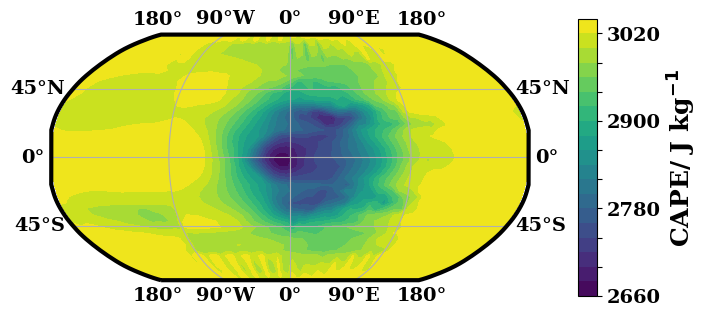}
    \caption{K2-18~b: average CAPE values in $J kg^{-1}$. They are higher than the TRAPPIST-1e equivalent values, reflecting the greater height of the convective region and the greater temperatures, even if the radiative forcing is weaker. There is much less geographical variation in CAPE, reflecting the role of convection in maintaining the globally uniform P-T profile in the deeper atmosphere. It is lower in the substellar region as a result of the general atmospheric overturning circulation.}
    \label{fig:k2-18b_CAPE}
\end{figure}

Global variations in Convective Available Potential Energy (CAPE) as shown in Figure \ref{fig:k2-18b_CAPE}. Convection is present globally, as expected from the generally globally uniform P-T profile. The CAPE values are higher than in the TRAPPIST-1e cases (Fig \ref{fig:CAPE_new}), a reflection of the greater pressure range convection acts over and of the higher temperatures it happens at, although we see relatively weaker convection at the substellar point. 

This is because of the global overturning circulation (seen in Figure \ref{fig:k2-18b_msf}): the general rising of the whole atmosphere here reduces the degree of instability to sub-grid parametrised convection, because the resolved bulk motions of the whole atmosphere already act to destroy CAPE. This reduction in convection at the substellar point is seen in other thick atmosphere cases such as \citet{Turbet2023} and \citet{Turbet2021}, and is related to where the stellar energy is deposited. 
We expect to see this less-frequent substellar convection in conditions with greater shortwave absorption across a range of the atmosphere, as opposed to scenarios where the bulk of incoming stellar energy is deposited at the surface (e.g. a thin $N_2$ dominated atmosphere as in our TRAPPIST-1e case). In the first case, heating will be fairly evenly spread across the troposphere and the general overturning circulation can destroy CAPE. In the second case, the majority of the stellar energy is deposited in the surface and so atmospheric surface layers.This means that the surface layers specifically become unstable in a way that bulk motions cannot address, leading to convection starting from the surface.

\begin{figure*}
    \centering
    \includegraphics[width=\textwidth]{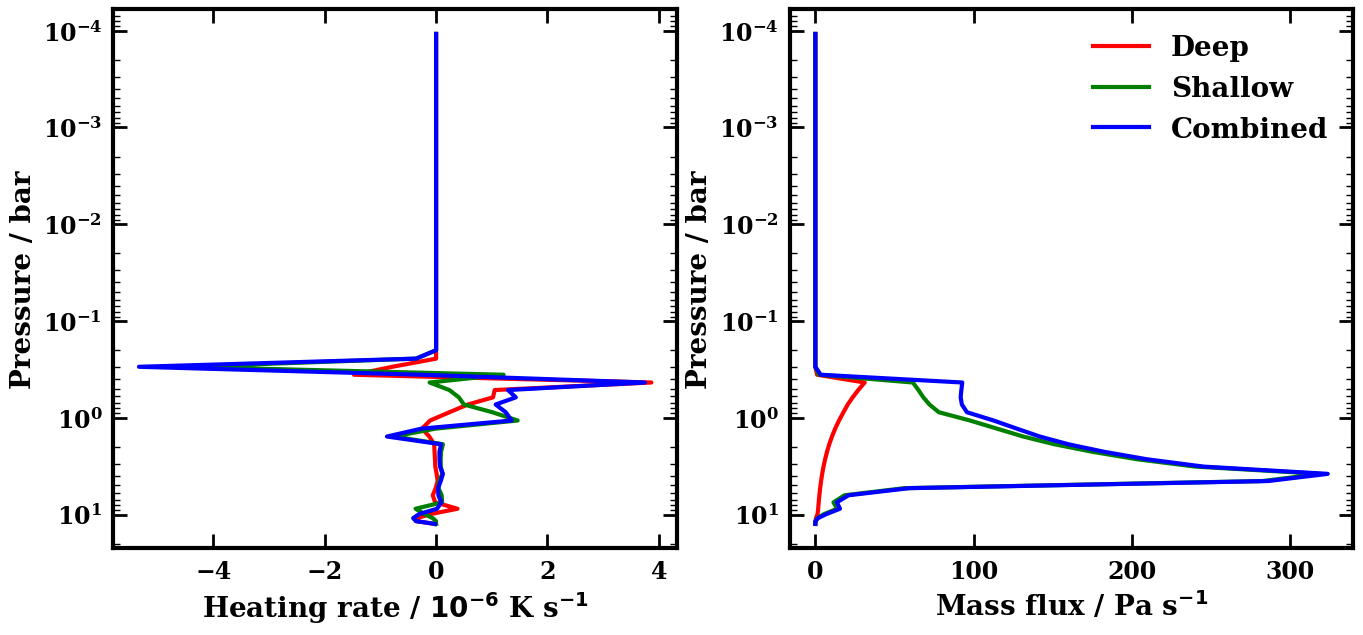}
    \caption{Left: The globally-averaged heating rates from deep and shallow convection in our K2-18~b run. Right: The average mass flux upwards (updraft - downdraft) from deep and shallow convection. The magnitude of the heating fluxes are similar from both deep and shallow convective schemes, and show an overall cooling-heating-cooling trend as expected, but there is a degree of messiness in the level-by-level heating rates. The deep convective mass flux steadily increases as the entrainment rate is always positive right up to the plume top. The shallow convective mass flux is higher than the deep one but peaks at a lower altitude, around the 3 bar level.}
    \label{fig:k2-18b_zmdtdqs}
\end{figure*}

The convective heating rates from both deep and shallow convection, along with the respective mass fluxes, are shown in Figure \ref{fig:k2-18b_zmdtdqs}. There are some important qualitative differences from the TRAPPIST-1e moist convection. The first is how the deep convective mass fluxes evolve. Because we have no condensation and so no downdrafts, the entraining updrafts grow exponentially until they lose buoyancy. Whether the convective mass flux profiles look like this is an open question - our evidence suggests that they do, but in the absence of 3D cloud-permitting modelling it is hard to say so confidently. The shallow convective mass fluxes are also generally much higher than the deep ones, especially in the lower atmosphere, peaking at over 300 $Pa$ $s^{-1}$ at around 4 bar, equivalent to roughly a tenth of the atmosphere at that point being in a convective upwelling zone.

This scale of shallow convection is interesting, especially given that the magnitude of the heating trends resulting from deep and shallow convection are similar. The latter is to be expected, given that by construction the deep convection scheme acts to eliminate 2/3rds of the calculated CAPE and the shallow convection scheme should eliminate the rest. A reason we see more shallow than deep convection is that deep convection can transfer heat over much greater distances. In cases where the virtual potential gradients are small and smoothly varying like we see here - but not in other cases with localised temperature and moisture forcing from radiation, evaporation, and condensation - the deep convection scheme will be able to transfer heat much more efficiently than the shallow one, and so require lower mass fluxes.

We also see a complex and somewhat intricate picture with the heating trends. Overall the expected pattern of convection is seen - there is cooling at the base of the convective zone, then heating, then cooling again. This matches what we would expect in an atmosphere with different radiative heating rates, with a uniform lapse rate imposed by convection. However, the heating rates in specific layers are more uneven than might be expected. This is mostly due to the fact that the convection scheme does not act naively and simply try to restore a convection region to overall marginal stability, but simulates a convective plume as it would actually rise, leading to over-correction in some areas and under-correction in other. A second contributing factor, on the other hand, is that the exponential growth in deep convective plume size leads to heating rates which are weighted towards the top of the plume, which may or may not happen in reality.

\section{Summary and Discussion}
\label{sec:6}

Recent observations have indicated a large number of planets located between Earth and Neptune in size. These planets are theorised to have diverse compositions - super-Earths, mini-Neptunes, waterworlds, Hycean planets, lava planets. Extracting information about their atmospheres from observations necessitates accurate modelling of planetary and particularly atmospheric processes, a key process of which is convection. Accurately modelling convection is a particularly pertinent problem for these planets because they are expected to have various modes of convection, different to that found on Earth. In a `non-dilute' atmosphere with a high condensible mass-mixing ratio $q$, convection can first enter a stronger but more time-variable regime, before convection becomes overall weaker and more sluggish. In a $H_2$ dominated atmosphere, moist convection can shut down entirely if the condensible exceeds a certain $q_{crit}$.

To enable accurate GCM simulations of these atmospheres, we develop an updated convection scheme that can accurately model the different modes of convection described above. The convection schemes we build upon are the \cite{Zhang1995} deep convection scheme and the \citet{Hack1993} shallow convection scheme, which are used in the ExoCAM GCM \citep{Wolf2022}. ExoCAM is an offshoot of the Community Earth Systems Model (CESM) GCM \citep{Neale2010} modified for exoplanet use.

Several important changes are made to the convection schemes. We evaluate stability to both dry (using virtual potential temperature) and moist convection using the criterion in Eq \ref{eq:Leconte17_15}. Deep convection starts at the lowest unstable layer, but can start higher up than that if this allows convection to happen. 
Updraft entrainment and detrainment rates, crucial parameters controlling the evolution of the plume and the resulting heating and drying rates, are calculated by using entropy conservation in the upwards plume. An entropy-based method is used to calculate the updraft and downdraft properties. We account for the mass of the updraft and downdrafts changing due to precipitation losses. A more accurate approach is used for the CAPE-based closure. We allow for the existence of two extended convective regions.
We modify the start criterion of the Hack shallow convection scheme, making it more flexible, and also allow for the existence of dry convection. 

We test the model on two cases. First, we test the performance of the scheme in conditions where the expected results are well known. We simulate the terrestrial exoplanet TRAPPIST-1e in an aquaplanet configuration with a $N_2$-CO$_2$ atmosphere. Comparing to published results (the Hab 1 case from the THAI collaboration, \cite{Fauchez2020}), the new scheme shows excellent agreement, showing that the model works in the heavy background atmosphere, low-$q$ limit. Deep convection is localised to the substellar region. The closer to the substellar point, the stronger and higher it goes, and the more heat and moisture is deposited into the upper troposphere.

We explore a GCM for a mini-neptune scenario, adopting the bulk properties and instellation of K2-18~b. We assume a 100x solar metallicity and fix the H$_2$O mass fraction at the model lower boundary.  
We find somewhat cooler temperature than the equivalent \citet{Charnay2021} circulation, a difference we attribute to the radiative scheme used. We find super-rotating tropospheric jets at mid-latitudes. Convection acts across the whole planet between 0.3 and 10 bar, but more weakly in the substellar region - a result of the global overturning circulation reducing the need for sub grid-scale convection there.

Overall, we have demonstrated the validation of a new deep convection scheme in a range of different scenarios, and which holds great promise for future studies of various exoplanet atmospheres, including future modelling of Hycean scenarios.

\subsection{Applicability of the new scheme}
The new scheme can capture a wide range of conditions where convection happens. It can handle moist or dry convection, rising from any point in the atmosphere, whereas the Zhang-Macfarlane and Hack schemes works best with moist convection, rising from near the surface.
It now handles compositional gradients (which can either facilitate or hinder convection), moist convective inhibition (in light background atmospheres), and convection in non-dilute atmospheres where the condensible can be a substantial fraction of the atmosphere by mass. It does this all whilst still modelling the mechanisms of convection and keeping state-of-the-art precision in the more familiar terrestrial exoplanet cases. This makes it an appealing option for general use in exoplanet GCMs.

There are, as always, some limitations in our scheme. For numerical reasons, there must be a dry gas component. It cannot handle atmospheres made uniquely of a condensible component and sitting along the saturation $p(T)$ profile. However, as long as there is an amount of non-condensible background gas the scheme works adequately. The transport of momentum by precipitation is also not included, which might impose limitations when this becomes large.

One important factor to be aware of relates to the pressure range covered by convection. On Earth and similar terrestrial planets, this is usually about half an order of magnitude (e.g. from 1 bar to $\approx$ 0.3 bar). In our mini-Neptune K2-18~b case presented in Section 5, we see convective plumes stretching over one and a half orders of magnitude. These are probably approaching the limits of how far our entrainment-detrainment rate calculation in section \ref{sec:3_2} can take us, without additional consideration of how plumes may exchange energy with their environment even if they are buoyant. For deeper atmospheres, such as GCMs run with model bottom pressures of 100 bar, it may be necessary to use insights from stellar and `abyssal' convection \citep{Markham2023} to further develop convection schemes. 

Furthermore, convection is not the only sub-grid process present in GCMs. Diffusion processes, particularly in the planetary boundary layer, are also important transports of heat, moisture and momentum. Particularly in cases where convection shuts down (e.g. the suggested stable layer in sub-Neptunes \cite{Leconte2024}), proper consideration of diffusion is needed for a wholistic understanding of the atmospheres.

\subsection{General applicability of mass-flux schemes}

There is also a legitimate discussion to be had about how far mass-flux schemes can be applied. Whilst more sophisticated and leading to more accurate heat and moisture transport and thermal profiles when compared to convective adjustment schemes, mass-flux schemes are themselves simplifications of reality. They do not capture every process happening in dry or moist convection. On Earth, whilst this is not completely solved (see the remaining IPCC uncertainties originating from convection-cloud coupling or the limited understanding of e.g. the Madden-Julian Oscillation), convection schemes still reproduce the modern climate to a high degree of accuracy. But how can we be confident in the various design choices and assumptions of the mass-flux scheme under these different conditions?

The first category of choices we can be confident in and we expect to be universal. The general theory of convection - starting from locally unstable layers, it restores a region of the system to roughly marginal stability - is sound. Similarly, the details of how a plume rises is driven by thermodynamics and are governed by the same laws everywhere. 
What is more complicated is knowing which of the key nuts and bolts of mass-flux schemes (updrafts, downdrafts, assumptions about which levels to start each from, which extra energy fluxes need to be included, the inclusion of deep and shallow convection schemes are valid in different modes of convection. Generally mass-flux schemes have these decisions made by observing the  details of moist convection as it occurs on Earth. As we have a limited ability to directly observe the mechanics of these different convective modes (and no way to do that for exoplanets), we cannot be as confident in our choices. That does not mean we cannot make any progress.

The first approach is to use observations of convection happening in solar system planets. Observations of storms and lightning in the atmosphere of Jupiter \citep{Gierasch2000} show that moist convection in a $H_2$-dominated atmosphere leads to the same kind of convective supercells, vigorous updrafts, and time variability as on Earth. Similarly, Venus Express pictures of convective cells on Venus \citep{SanchezLavega2017} show convective cells of the sort familiar on Earth.

The other route we can use to gain insight about the mechanics of convection is high-resolution simulations i.e. CPMs. With horizontal resolutions in the hundreds to thousands of metres, we can resolve convection explicitly. 
These CPMs give a largely reassuring picture - the fundamentals of convection are intact. In the terrestrial exoplanet case \citep{Sergeev2020}, convection works largely like on Earth, as expected given the atmospheric conditions. We do see some interesting effects tied to the presence of a global ocean or possibly constant dayside heating, notably the development of banded cloud formations.

\citet{Habib2024a} performed CPM simulations of non-condensing compositional convection. Although this study focuses on an initial instability mixing to equilibrium and does not consider what the quasi-steady state in the presence of radiative heating would look like, it still provides useful information. We see mixing once the plume reaches marginal stability with the environment. There does not appear to be much entrainment, although the large size of the updrafts relative to the environment and the height they rise may preclude this. Updrafts are larger and more numerous than downdrafts, justifying the focus on including and modelling them. One important result is frequent, but not universal, formation of 'compositional staircases'. These staircases can occasionally stop the initial unstable layer fully mixing, and may need to be considered in parametrisations.

The \citet{Leconte2024} results are interesting because they do show a very different convection structure happening in two layers. In the lower section of the modelled atmosphere, there is dry convection, and then moist convection occurs above a stable layer as predicted in \cite{Leconte2017}. Occasional plumes graze into the stable layer, showing the importance of allowing some degree of convective overshooting. The existence of a small dry boundary layer between the stable layer and the moist troposphere above is reassuring, as it mirrors the dry surface boundary layer we see on Earth. In both cases, small scale dry convection transports plumes upwards until they saturate and moist convection begins. That this feature is the same in these two very different types of moist convection means that the underlying mechanics should be similar. We also see some downdrafts on the edges of the updraft in the moist troposphere. Something more challenging to simulate is the presence of a convective downdraft driven by evaporating rainwater \textit{below} the stable layer, in the dry convective region. This shows the necessity of coupling different convective regions in a single column.

\citet{Habib2024b} and \citet{Seeley2024} also looked at moist convective inhibition in a hydrogen-dominated atmosphere. They found a broadly similar picture: as expected from linear theory, a stable layer develops where convection is inhibited if $q>q_{crit}$. If the model bottom is fixed to be subsaturated, then the most common result is for a dry convective layer, then the stable layer, then some moist convection above. If the bottom is saturated, as it would be for an atmosphere in contact with a liquid water surface, the stable layer starts at the surface. 

In the \citet{Habib2024b} work, it is difficult to draw strong conclusions about the final steady state of the system as the simulations have not been run to a statistical-equilibrium state. However, there are indications that diffusion may be weaker than in the \cite{Leconte2024} work, as latent heat transport is smaller by an order of magnitude, and the exact details of the moist physics microphysics parametrisations used also seem to make a difference. Conversely, the \citet{Seeley2024} results, which use a simplified radiative transfer model to focus in on the convective dynamics, appear to find that the near-surface stable layers have higher lapse rates than free tropospheric stable layers, and also see some trace of emergent periodicity in convective activity. That these separate pieces of work have all contributed valuable insights into moist convective inhibition shows the value of comparative studies and of using a series of different models to examine the same problem.

The results of \citet{Seeley2023} also suggests that mass-flux schemes remain widely applicable. They show a continuity of convective behaviour from Earth-like conditions to non-dilute atmospheres. Convective vigour peaks at a surface temperature of about $330K$ (for a 1 bar atmosphere), but the general features of convection remain the same throughout a wide range of surface temperatures, including a predominance of updrafts, compensating subsidence, and convective overshooting into stable layers.

Overall there are some differences in the puzzle pieces needed to accurately simulate convection in different regimes, notably in the role and nature of downdrafts. However, the available evidence suggests that the basic schematic used for Earth-like convection can be extended to other modes of convection.

\subsection{Comparison with previous GCMs of mini-Neptune scenario}
Two other GCM studies of K2-18~b are \cite{Charnay2021} (henceforth referred to as C21) and \cite{Innes2022} (I22). Our results, presented in Section \ref{sec:5}, present many similarities but also some important differences to these two works.

A first key similarity is the low level of global temperature variation except in the top layers of the atmosphere, where there is a dayside temperature inversion. This matches both C21 and I22 and confirms that we are in a `WTG' regime. The vertical temperature profile is also broadly similar to C21, showing a minimum in the temperature around the $10^{-2}$ bar region. At higher pressures, we find lower temperatures as the near-isothermal zone extends further, although we do not find the drop in lapse rate towards the model bottom. These differences are all reflections of the ExoRT radiative transfer scheme used \citep{Wolf2022}: a 1D version of the radiative transfer scheme gives a very similar P-T profile to the one shown in \ref{fig:k2-18b_selected_PTs}. We note that the ExoRT cross sections are only accurate to 500K and so, while the temperature impact on the opacities is much lower than that of e.g. the abundances, the P-T profile of the deep atmosphere may have some inaccuracies.

Even relatively small temperature differences can have significant impacts on other aspects of the atmosphere - here, notably, a difference of less than 10K around the $10^{-2}$ bar level means that our atmosphere is almost completely cloud-free whereas the C21 100x metallicity run has significant clouds. The C21 P-T profile does not intersect the dewpoint temperature profile by much (although this is partly driven by the release of latent heat which warms the atmosphere in those layers). This is a significant difference with implications for observations, and further illustrates the impact of differences in radiative transfer schemes.

The dynamical structure also presents interesting contrasts to existing results. Despite the inclusion of surface drag (which leads to the atmosphere angular momentum not being conserved over time and to some atmospheric super-rotation) matching C21 and not I22, our results are overall more similar to the later.
We also find high-latitude super-rotating jets, and equatorial super-rotating jets nearer the model top, similar to Figure 5 of I22. Our bottom layers have a positive velocity in contrast to I22, which is likely a result of the surface drag imposed. We note that such a stratospheric equatorial jet cannot develop from a stationary planetary wave as the dimensionless Rossby radius is too high, and that some other mechanism such as barotropic waves is needed to generate the stratospheric super-rotation seen. However, our tropospheric super-rotating jet could plausibly be formed by such a mechanism. Also similarly to I22, we find that the dominant overturning circulation is day-night but that the pole-equator overturning circulation is also substantial. Using a Helmholtz decomposition \cite{Hammond2021}, we also find that the velocity field is dominated by the rotational circulation and not the divergent circulation. That said, we find that the divergent circulation is responsible for the vast majority of the day-night energy transport, despite its much lower amplitude.

These results can be contrasted with C21, who found a weak equatorial jet at the 0.1 - 10 bar level, retrograde jets at higher altitude and latitudes, and a dominant overturning circulation. The reason for this difference is not clear - we also impose a surface drag, have a real-gas radiation scheme, and work on a latitude-longitude grid, like C21, yet our dynamical structure is more like that found in I22. The likelihood is that some other detail of the dynamical core drives these differences - the model base pressure being at 80 bar instead of 10 bar could feasibly make an important difference, or the difference in radiative temperature gradients leads to these differences. We also note that of the C21 runs with different parameters have significantly different zonal wind structures, so it is likely that small changes in the parameters such as the atmospheric composition could dramatically influence the zonal wind structure.

\section*{Acknowledgements}

 E.B. and N.M. acknowledge support from the UK Research and Innovation (UKRI) Frontier Grant (EP/X025179/1, PI: N. Madhusudhan) towards the doctoral studies of E.B. This work was performed using resources provided by the Cambridge Service for Data Driven Discovery (CSD3) operated by the University of Cambridge Research Computing Service (www.csd3.cam.ac.uk), provided by Dell EMC and Intel.

\section*{Data Availability}

 The default ExoCAM GCM which this convection code is built off is available at https://github.com/storyofthewolf/ExoCAM. Models outputs are available on request from the paper authors.



\bibliographystyle{mnras}
\bibliography{paper} 




\appendix

\section{Condensation and precipitation in the updraft}
\label{sec:a1}

This section covers how the condensation and precipitation rates are calculated, and how we carry out the mass scaling of the updraft plume to account for precipitation losses. It is analogous to equations 4.25-4.26 and 4.71-4.77 of \cite{Neale2010}.

We start with the same governing equation for plume liquid MMR as 4.25

\begin{equation}
    \frac{\partial}{\partial z} (M_u l) = - D_u l_D - R_u + C_u
    \label{eq:ql_governing}
\end{equation}

With $M_u$ the updraft mass, $l$ the liquid water MMR, $D_u$ the detraining rate, $l_D$ the detraining liquid water MMR, $C_u$ the condensation rate and $R_u$ the precipitation rate. Liquid cloud water autoconverts to rainwater at a fixed rate of $c_0 = 2\times10^{-3} m^{-1}$ \citep{Lord1982}, and we write the precipitation rate $R_u$ as 

\begin{equation}
    R_u = c_0 M_u l
    \label{eq:Ru}
\end{equation}

To find the condensation rate $C_u$, we use $l_{in}$ and $l_{out}$ corresponding to the liquid water MMR before condensation, and after condensation but before autoconversion to precipitation and any mass scaling. $l_{out}$ is calculated from conserving entropy as described in Section 3.3 and $l_{in}$ is calculated similarly to Equation \ref{eq:hu_updraft_calc}.

\begin{equation}
    l_{in} = \frac{M_u^{k+}}{M_u^{k-}} l^{k+} - \frac{dz^k}{M_u^{k-}} D_u^k l_D^{k+}
    \label{qlin}
\end{equation}

\begin{equation}
    C_u = (l_{out} - l_{in}) \frac{M_u^{k-}}{dz^k}
    \label{eq:Cu}
\end{equation}

Differencing Equation \ref{eq:ql_governing}, we substitute in \ref{eq:Cu} and \ref{eq:Ru} to find

\begin{equation} \begin{split}
    \frac{M_u^{k-} l^{k-} - M_u^{k+} l^{k+}}{dz^k} =& - D_u^k l_D^{k+} - c_0  M_u^{k-} l^{k-} + \\& \frac{M_u^{k-}}{dz^k} (l_{out} - \frac{M_u^{k+}} {M_u^{k-}} l^{k+} - \frac{dz^k}{M_u^{k-}} D_u^k l_D^{k+}) 
\end{split} \end{equation}

Cancelling, this gives us our final expression for the liquid water MMR (before the mass-scaling):

\begin{equation}
    l^{k-} = \frac{l_{out}}{1 + c_0 dz^k}
\end{equation}

\section{CAPE}
\label{sec:a2}
The Convective Available Potential Energy (CAPE) is the energy released as the parcel moves from its starting position to its level of neutral buoyancy (LNB), where it finishes its ascent (give or take some convective overshooting). We adopt the usual expression

\begin{equation}
    CAPE = R^d \int_{LNB}^{p_{bot}} (T_{p,v} - T_{e,v})  d \ln p
\label{eq:CAPE}
\end{equation} 

One of the main design choices made by convection schemes is the type of closure used to calculate the total mass fluxes. The ZM scheme uses a CAPE based closure, destroying available CAPE at a fixed rate. We keep this, but the CAPE closure equation used is slightly modified. Given the CAPE equation and denoting the term $T_{p,v} - T_{e,v}$ as $B$, 

\begin{equation}
    \frac{dB}{dt} = \frac{d}{dt} [T_p(1 - \bar{\omega} q_p) - T_e(1 - \bar{\omega} q_e)]
\end{equation}

The rates of changes of $T_e$ and $q_e$ can be calculated from considering the plume mass fluxes, temperatures and humidities as outlined in \citet{Neale2010}. To find $dT_p/dt$ and $dq_p/dt$, we take the changes in $T$ and $q$ in the sub-convective region $dT_b/dt$ and $dq_b/dt$, and by tracing an entraining upwards plume find the values of $dT_p/dT_b$, $dT_p/dq_b$, $dq_p/dT_b$, and $dq_p/dq_b$. The relevant derivatives are then calculated:

\begin{equation}
    \frac{dT_p}{dt} = \frac{dT_p}{dT_b} \frac{dT_b}{dt} + \frac{dT_p}{dq_b} \frac{dq_b}{dt}
\end{equation}

\section{Dry convection in the Hack scheme}
\label{sec:a3}

The standard Hack scheme only acts when a parcel from a given layer would reach the LCL and be buoyant when moved to the layer above, i.e. when $h^{k+1} > h_{sat}^k$. We allow convection if the parcel originating from $k+1$ is buoyant at level $k$. If this happens and there is condensation, we treat it as moist convection following \cite{Neale2010} Equations 4.86-4.113. The equations for the dry are similar - effectively, we set the liquid water MMR $ql^k$ to zero (alongside linked quantities like the rain production rate $R^k$), and replace the moist static energy by the dry static energy where appropriate. The conserved quantity is the dry static energy $s$ and not the 'liquid water static energy' $s_l = s - L$  $ql$. The governing equations (\cite{Neale2010} 4.91-4.96) become

\begin{equation}
    \hat{s}_{k-1} = s_{k-1} + \frac{2\Delta tg}{\Delta p_{k-1}} (\beta m_u (s_u - s_{k-\frac{1}{2}}))
\end{equation}
\begin{equation}
    \hat{s}_{k} = s_{k} + \frac{2\Delta tg}{\Delta p_{k}} (m_u (s_u - s_{k+\frac{1}{2}}) - \beta m_u (s_u - s_{k-\frac{1}{2}}))
\end{equation}
\begin{equation}
    \hat{s}_{k+1} = s_{k+1} + \frac{2\Delta tg}{\Delta p_{k+1}} (m_u (s_{k+\frac{1}{2}} - s_u))
\end{equation}

\begin{equation}
    \hat{q}_{k-1} = q_{k-1} + \frac{2\Delta tg}{\Delta p_{k-1}} (\beta m_u (q_u - q_{k-\frac{1}{2}}))
\end{equation}
\begin{equation}
    \hat{q}_{k} = q_{k} + \frac{2\Delta tg}{\Delta p_{k}} (m_u (q_u - q_{k+\frac{1}{2}}) - \beta m_u (q_u - q_{k-\frac{1}{2}}))
\end{equation}
\begin{equation}
   \hat{q}_{k+1} = q_{k+1} + \frac{2\Delta tg}{\Delta p_{k+1}} (m_u (q_{k+\frac{1}{2}} - q_u))
\end{equation}

Where the subscript $_u$ denotes updraft quantities, $m_u$ is the convective mass flux at the bottom of the middle layer (level $k + \frac{1}{2}$, cloud base), and $\beta$ is the (so far unfixed) detrainment parameter which controls what fraction of the $m_u$ will penetrate into the $k-1$ layer. $\Delta p$ refers to the pressure thickness of a layer, $g$ is the surface gravity, and $\Delta t$ is the duration of the model timestep. We can write the mass flux as (following 4.108)

\begin{equation}
    m_u = \frac{s_u - s_k}{g\tau (\frac{1}{\Delta p_k}(s_u - s_{k+\frac{1}{2}} - \beta (s_u - s_{k-\frac{1}{2}})) - \frac{1}{\Delta p_{k+1}} (s_{k+\frac{1}{2}} - s_u)}
\end{equation}

With $\tau$ a characteristic adjustment timescale. To find $\beta$, the first constraint - that the convective mass flux $m_u$ be positive - leads to (following 4.109)

\begin{equation}
    \beta (s_u - s_{k-\frac{1}{2}}) < (s_u - s_{k+\frac{1}{2}}) (1 + \frac{\Delta p_k}{\Delta p_{k+1}})
\end{equation}

In the moist convective case there is a second constraint preventing supersaturation of the detrainment layer at $k-1$. Because we are far from saturation, we do not impose this constraint. We do impose the third constraint, designed to prevent the injection of thermodynamic noise into the temperature profile (following 4.111, still from \cite{Neale2010}, and with $G$ an arbitrary vertical difference in the adjusted profile following \citet{Hack1993}):

\begin{equation} \begin{split}
    & \beta [\frac{s_k - s_{k-1} -G}{s_u - s_k} \frac{\tau}{2\Delta t} \frac{1}{\Delta p_k}(s_u - s_{k-\frac{1}{2}}) + (\frac{1}{\Delta p_k} + \frac{1}{\Delta p_{k-1}})(s_u - s_{k-\frac{1}{2}})] \leq \\& (\frac{s_k - s_{k-1} -G}{s_u - s_k} \frac{\tau}{2\Delta t} (\frac{1}{\Delta p_k} + \frac{1}{\Delta p_{k+1}})(s_u - s_{k+\frac{1}{2}})) - \frac{1}{\Delta p_k} (s_u - s_{k+\frac{1}{2}})
\end{split} \end{equation}

\bsp	
\label{lastpage}
\end{document}